# Microcircuit synchronization and heavy tailed synaptic weight distribution in preBötzinger Complex contribute to generation of breathing rhythm

Valentin M. Slepukhin[a]*, Sufyan Ashhad[b]*, Jack L. Feldman[b]§, and Alex J. Levine[a,c,d]§.

[a] Department of Physics and Astronomy, UCLA, Los Angeles California, 90095-1596, USA
[b] Systems Neurobiology Laboratory, Department of Neurobiology, David Geffen School of Medicine, University of California Los Angeles, Los Angeles,
California 90095-1763, USA
[c] Department of Chemistry and Biochemistry, UCLA, Los Angeles California, 90095-1596, USA
[d] Department of Computational Medicine, UCLA, Los Angeles California, 90095-1596, USA
*These authors contributed equally to this project
§These authors jointly supervised this project

**ABSTRACT**

The preBötzinger Complex, the mammalian inspiratory rhythm generator, encodes inspiratory time as motor pattern. Spike synchronization throughout this sparsely connected network generates inspiratory bursts albeit with variable latencies after preinspiratory activity onset in each breathing cycle. Using preBötC rhythmogenic microcircuit minimal models, we examined the variability in probability and latency to burst, mimicking experiments. Among various physiologically plausible graphs of 1000 point neurons with experimentally determined neuronal and synaptic parameters, directed Erdős-Rényi graphs best captured the experimentally observed dynamics. Mechanistically, preBötC (de)synchronization and oscillatory dynamics are regulated by the efferent connectivity of spiking neurons that gates the amplification of modest preinspiratory activity through input convergence. Furthermore, to replicate experiments, a lognormal distribution of synaptic weights was necessary to augment the efficacy of convergent coincident inputs. These mechanisms enable exceptionally robust yet flexible preBötC attractor dynamics that, we postulate, represent universal temporal-processing and decision-making computational motifs throughout the brain.

**INTRODUCTION**

The preBötzinger Complex (preBötC) is the kernel of the central pattern generator for breathing that produces the inspiratory rhythm in mammals (Del Negro et al., 2018; Smith et al., 1991), for which emergent network mechanisms, in addition to synaptic and cellular properties, play an essential role (Ashhad and Feldman, 2020; Cui et al., 2016; Del Negro et al., 2018; Feldman and Kam, 2015; Kam et al., 2013a; Pace et al., 2007a, b; Wang et al., 2014). Inspiratory motor output in each breathing cycle emerges only when putatively rhythmogenic (Ashhad and Feldman, 2020; Del Negro et al., 2018; Feldman and Kam, 2015; Gray et al., 1999; Kam et al., 2013a) Type I preBötC neurons, initially firing at low frequency in the preinspiratory period (preI), progressively synchronize to drive bursts of action potentials (APs) in Type II preBötC output neurons (Ashhad and Feldman, 2020; Cui et al., 2016; Gray et al., 1999; Kam et al., 2013a; Sun et al., 2019; Tan et al., 2008) that ultimately drive inspiratory motor activity (Figure 1A-D). Thus, the evolving activity of Type I neurons sets the onset time of each inspiration. Strikingly, preI, which marks the time for initial synchronization of Type 1 neurons, varies considerably from cycle to cycle and across experiments: 240±10 ms, mean±SEM *in vitro* (Ashhad and Feldman, 2020; Baertsch et al., 2018; Carroll and Ramirez, 2013; Del Negro et al., 2018; Gray et al., 1999; Kam et al., 2013a) (Figure 1 B-C); 160±32 ms, mean±SEM in vivo (Cui et al., 2016; Guyenet and Wang, 2001). This suggests that the process of preBötC network assembly leading to inspiratory burst generation is highly variable and emerges from dynamical cell assemblies in every cycle (Ashhad and Feldman, 2020). Furthermore, since the putative rhythmogenic Type I neurons appear sparsely (~13%) connected (Ashhad and Feldman, 2020; Rekling et al., 2000), a conundrum arises as to how synchrony emerges from a low level of activity. Here, we developed models of preBötC rhythmogenic microcircuits constrained by experimentally derived neuronal and synaptic parameters to better understand the network dynamics and the underlying network topology that gives rise to the rhythmicity of breathing.

We modeled the preBötC by treating its constituent neurons as low dimensional, nonlinear dynamical systems interacting on a quenched (and frequently random) directed graph, an approach pioneered in other neural networks (Ermentrout, 1998; Gal et al., 2019; Gerstner, 1995; Perin et al., 2011). In such studies, neuronal models fall into two distinct classes: (i) *firing rate* models that treat neuronal output as a single firing rate variable but ignore the temporal structure of the underlying spike trains, or; (ii) *spiking* models that consider the timing of each spike in the network where neuronal interactions depend on the temporal coincidence of discrete excitatory postsynaptic potentials (EPSPs) rather than on rate-based highly smoothed temporal summation (Bernander et al., 1994; Diesmann et al., 1999; Kumar et al., 2010; Lindsey et al., 2000; Wang and Buzsaki, 1996).

Most models of the preBötC are based on firing-rate dynamics, e.g., (Bacak et al., 2016; Schwab et al., 2010). These models account for three dynamical states: (i) a state with alternating periods of global activity and quiescence as in normal breathing; (ii) a permanently quiescent state, i.e., apnea, and; (iii) a state characterized by chaotic inspiratory bursts, as seen in many instances of brainstem pathology (Bacak et al., 2016; Bibireata et al., 2020; Del Negro et al., 2002). However, these models do not incorporate recent findings that suggest that preBötC inspiratory burst initiation in each cycle is a consequence of progressive synchronization of neurons comprising the rhythmogenic microcircuit (Ashhad and Feldman, 2020), nor can they account for the observation that the simultaneous stimulation of a small subset of neurons (4-9) can induce a global response, i.e., a network burst, at considerable delay, ~60-400 ms *in vitro* (Kam et al., 2013b). Moreover, these models do not consider evidence that there are at least two distinct subpopulations of excitatory neurons, one rhythmogenic that projects to a second population of output neurons generating and transmitting inspiratory bursting from the preBötC to efferent targets, e.g., inspiratory premotoneurons in the ventral respiratory column, but are not themselves rhythmogenic (Ashhad and Feldman, 2020; Cui et al., 2016; Del Negro et al., 2018; Feldman and Kam, 2015; Kam et al., 2013a; Tan et al., 2008; Yang and Feldman, 2018).

In each breathing cycle (starting at the end of the previous inspiratory burst), there emerges at a delay low-frequency synchronized firing of neurons in the preBötC rhythmogenic microcircuit, i.e., a *burstlet* (Ashhad and Feldman, 2020; Kam et al., 2013a; Sun et al., 2019) that precedes and ultimately triggers, with high probability, each inspiratory burst (Figure 1 B). The dynamics of network synchronization underlying burstlets and subsequent burst formation is the focus of this work. We do not consider the termination of network bursts or the transfer of bursts to the output targets of the preBötC.

In *in vitro* experiments in inspiratory rhythmic slices, simultaneous activation of a small subset of inspiratory-modulated neurons (4-9) can initiate a burstlet at significant delay (170 ms - 370 ms) with >80% reliability (Kam et al., 2013b) (Figure 1E-H). We sought to determine network features required to simulate burst formation dynamics consistent with the experimentally observed network connectivity and synchronization, including the significant delay between the initiation of induced firing of a few neurons and the globally synchronized firing of the network of ~1000 neurons, i.e., bursts (Figure 1 G, H). In these experiments, while the mean delay time decreases reproducibly when the number of initially activated neurons increases, the latency to burst varies: (i) from trial to trial when the same neurons are stimulated, but in random order, and; (ii) when different randomly selected subsets of the same number of neurons are stimulated.

We emphasize that there is scant data on the connectivity within the preBötC, reflecting a large measure of uncertainty of its actual connectome. By exploring burst formation dynamics over a variety of plausible network connectivity schemes, we identified those networks consistent with experimentally observed dynamics. Using our model, we explored the propagation of spike synchronization across networks of different topologies (Figure 2 A, E, I, M), showing that only a small subset of connectomically reasonable network structures produced synchronization dynamics consistent with experimental observations. Importantly, we found the experimentally observed lognormal distribution of synaptic weights, as also found in other neuronal networks (Buzsaki and Mizuseki, 2014; Lefort et al., 2009; Loewenstein et al., 2011; Manabe et al., 1992; Song et al., 2005), appears crucial for network synchronization.

To understand the role of network connectivity on synchronization dynamics, we employed a combination of novel graph theoretic measures of network connectivity and activity with machine learning. For example, networks of directed Erdős-Rényi (ER) graphs with lognormal synaptic weights contained groups of neurons that *sent* greater weight and/or more efferent synapses (we call these neurons *high senders*), and other groups of neurons that *received* greater weights and/or more connections (*high receivers*) than the population average. If on a trial the initially activated neurons were high senders that projected to neurons that were also high senders, neuronal synchrony would likely emerge through input convergence and coincidence detection. Conversely, if the secondarily activated neurons were not high senders, the activity faded away. Thus, in this circuit, dynamic graph motifs encompassing efferent connectivity of spiking neurons can interact with ongoing network dynamics to generate large-scale network synchronization. We found that, amongst the physiologically reasonable graphs studied, only Erdős-Rényi (ER) graphs with lognormal distributions of synaptic weights produced synchronization dynamics consistent with experimental data. To better understand the graph motifs responsible for synchronization, we turned to machine learning to assess the relative importance of topological properties of the graph in producing global synchronization. We attribute the above observations to the nonlinear nature of spike coincidence detection which depends on the fat tail of synaptic weights and is gated by small subgraphs of the network that determine input convergence.

Thus, our study suggests an essential physiological role of a fat tail synaptic weight distribution underlying network computation leading to burst generation in the preBötC. Given the universality of lognormally distributed synaptic strengths across various brain regions (Buzsaki and Mizuseki, 2014; Schroter et al., 2017) our analysis reveals a generalized mechanism through which local topology dynamically interacts with the instantaneous network dynamics to regulate signal processing underlying a vital brain function. Finally, these mechanisms underlying network assembly for production of the inspiratory burst may represent a template for similar signal processing underlying other brain functions, such as in temporal processing and decision making (Breakspear, 2017; Buzsaki and Draguhn, 2004; Buzsaki and Mizuseki, 2014; Knierim and Zhang, 2012; Paton and Buonomano, 2018; Wang, 2010; Wimmer et al., 2014).

**Results**

**Randomly connected networks of excitatory neurons replicated preBötC synchronization**

We explored various models to simulate experimental stimulation of 4-9 preBötC neurons that induce inspiratory bursts (I-bursts; Figures 1, S1; also see METHODS) and examined the underlying network dynamics. Almost all network topologies with physiologically determined parameters we tested (see METHODS) synchronized at various latencies (Figure 2). Here, synchronization refers to the emergence of significant spike-to-spike correlated activity between and among a large fraction of neurons (Kumar et al., 2010; Ratte et al., 2013; Wang, 2010) that is a hallmark of preBötC inspiratory activity (Ashhad and Feldman, 2020) (Figure 3). Notably, no tuning of experimentally measured parameters was necessary. For both localized (Figure 2 A-D) and hierarchical (Figure 2 E-H) networks, which had significant clustering (a measure of how strongly neurons are interconnected into groups, see METHODS), the threshold number of stimulated neurons to synchronize the network with $\gtrsim$80% success rate ranged from 2-4 for majority of the networks tested over multiple realizations (each realization representing a different network but with similar topological properties) (Watts and Strogatz, 1998). We considered these networks too excitable, since the minimum observed experimentally was $\gtrapprox$4 stimulated neurons. Only randomly connected ER graphs could produce preBötC bursts with synchronization at a $\gtrsim$80% success rate in response to activating 4-9 neurons (Figure 2I-K) at latencies comparable to those seen experimentally (Figure 2 L, grey box). In these model networks, the outward connection from any neuron to any other neuron was based on the experimentally measured probability of 0.065 and included an additional random factor in connection strength in order to reflect the experimentally determined variance (Rekling et al., 2000). Consequently, the threshold number of stimulated neurons required to reliably ($\gtrsim$80%) synchronize a network varied among networks with the same overall parameters (see Figure 2K for the probabilities to synchronize; each curve is a different network with different choice of activated neurons). Illustrative of the effects of spike timing, Figures 2 H, D, L, and P show the variability in time to synchronize (as error bars) within a given network when the same neurons were activated in all simulations across various trials, *but in random order*. The range of mean latencies to synchronize of different ER networks was 39 ms – 235 ms, inversely related to the number of stimulated neurons (Figure 2L), overlapping with experiments (*cf*. Figure 2L with Figure 1G; mean latency for the threshold number of stimulated neurons: model: 73±3 ms – 184±15 ms, mean±SEM, across different network iterations; experiment: 255±43 ms (Kam et al., 2013b) mean ± SEM, range 170 ms – 370 ms).

Interestingly, when we tweaked the connectivity in these ER networks to increase the number of directed triangular edges, i.e., 3-simplices (see METHODS), by ~11% (with an expected shift in network topology to small world-like (Watts and Strogatz, 1998) (Figures 2 M-N; also see METHODS), the threshold number of neurons required to reliably synchronize the network decreased to 2 – 4 in 5 of 10 different realizations (Watts and Strogatz, 1998) (Figure 2 O) with decreased latency to synchronize (27±0.6 ms – 116±16 ms, mean±SEM) (Figure 2 P). Thus, an increase in the clustering coefficients led to a decrease in the number of neurons required for synchronization of the entire network (*cf.* Figures 2 B, F, J, and N).

**ER graphs with lognormal synaptic weights reproduced trial variability in latency to preBötC synchronization seen in experiments**.
We analyzed the dynamics of preBötC synchronization on ER networks (see below) by activating 7 randomly selected neurons (the sweet spot, in the range 4-9, for burst activation in experiments) over five trials (Figure 3, R1-R5). The key observation was that the time course of the activation pattern replicated essential features of the experimental data. In this network (as well as others summarized in Figures 2K-L), there was considerable variability in the evolution of activity associated with just small variations in the order of and precise individual activation times (±5 ms, SD) of the same subset of neurons. Bursts were not seen in every run (1 failure/5 runs) and when bursts were produced (4/5) there was considerable dispersion of delay times (99 ms - 195 ms; 145 ± 21 ms mean ± SEM) in the same range as experimental data with similar suprathreshold number of stimulated neurons (57 ms – 160 ms; 125 ± 23 ms; mean ± SEM) ( Kam et al., 2013b) (*cf*. Figures 3 A - C with Figure 1I)

Given that this model simulated experimental data, we explored the mechanistic underpinnings of preBötC synchrony and the variability in time to synchronize. The growth of network activity and the process of synchronization can be divided into two epochs. During the first epoch, the growth of activity was insignificant (Figure 3C), as the activity of the stimulated neurons was insufficient to initiate a burst. In fact, in some cycles, spikes from activated neurons did not induce any substantial spiking in their downstream (postsynaptic) neurons, e.g., first set of spikes for R1-R3 in Figure 3A. This could be due to the jitter in their spike times, insofar as the spikes elicited in the stimulated neurons were not sufficiently coincident to significantly activate, i.e., produce spikes, in their postsynaptic neurons. Consequently, the network activity died down to baseline once these neurons stopped firing in response to their initial stimulus (Figure 3B-C first set of spikes between 0-50 ms). With repeated activation of the stimulated neurons, due to both inherent jitter in the activation times and in the spiking of the activated neurons, their spikes bore higher coincidence in certain intervals. Consequently, the temporal summation of potentials in their postsynaptic neurons led to an activation of a critical group of neurons sufficient for synchronization, which initiated a second epoch. During this second epoch, this process of transient network activity continued until the stimulated neurons successfully induced spreading activity leading to increasing synchronization that fueled a rapid and sudden amplification of network activity. At this point, the simulations were terminated. In test cases, where simulation was allowed to continue, the network continued to fire in a prolonged synchronized bursting mode where the activity was modulated by the neuronal refractory period of 3 ms (Figure S3). For a run that did not end with network synchronization (Figure 3 A-C, R2), the group of neurons sufficient for synchronization was never activated, the simulation trial ended with the last spike of initially activated neurons (Figure 3A) and quenching of the whole system (Figure 3 B-C, red trial).

**Lognormally distributed synaptic weights enhanced fidelity of spike transmission, enabling network synchronization with fewer inputs**

Is the reliability of the neural networks on ER graphs in replicating experimental results solely an intrinsic property of the ER graph, or are the experimentally determined synaptic weights an essential feature of the network dynamics? To address this, we constructed ER networks with uniform weights equal to the mean synaptic weight as determined experimentally and compared them to those using a lognormal distribution for synaptic weights (Figure 4). All networks with uniform distributions (n=5) required stimulation of ≳15 neurons to initiate bursts with ≳80% success (*cf.* Figure 4A and B with Figures 2 K and L respectively). Strikingly, networks with uniform weights required more stimulated neurons to initiate bursts even when most of the weights in each lognormal network (61% in network of Figure 3A) were less than the distribution mean, which was also the weight of the uniform networks. This suggests that the average connectivity strength of the network is not a critical determinant for synchrony. We reason that the fat tail of the lognormal distribution allows neurons to have, with significantly high probability, synaptic weights that are significantly stronger than average, i.e., with ~ 5-10mV EPSP amplitude; Figure 4G, making the recipient neuron more likely to reach spike threshold for (near) coincident EPSPs (Figure S1) arising from synapses of high and low weight inputs, i.e., the networks with such connectivity paradigm are *better coincidence detectors* (Figure 4C-D). Here, coincidence detection is implied phenomenologically (Lisman, 1997), not accounting for various other neuronal and synaptic mechanisms contributing to intrinsic neuronal coincidence detection properties (Abbott and Nelson, 2000; Egger et al., 1999; Song et al., 2000).

We further explored the emergent neuronal signal processing properties arising from a lognormal distribution by simulating the behavior of a neuron with 50 inputs whose input synaptic weights were either lognormally or uniformly distributed. We analyzed 10 such all-to-one networks where 20% (10/50 randomly selected) of the input neurons were activated at various Poisson-distributed frequencies; this ensured that a broad range of synaptic weights in the lognormal distribution were activated (Figure 4E-G). For mean synaptic stimulation frequencies 10 - 200 Hz, neurons with lognormal inputs had a greater dynamic range of output frequencies (Figure 4E) and spike probability, i.e., total output spikes/total input spikes, (Figure 4F) as compared to neurons with uniform inputs. Their larger dynamic ranges indicate that a neuron with lognormal inputs can better *differentiate* among its presynaptic partners for the same input statistics, i.e., synaptic input frequency, as well as being better coincidence detectors, by virtue of the fat tail, rather than one with uniformly, distributed inputs. Notably, this asymmetry ensures high response reliability for all neurons, where higher synaptic strength of some partners ensures that most neurons can reliably respond to its presynaptic partners even for low frequency inputs, but only when strongly connected partners are coactive. This is reflected in upper tail of their response distributions (Figure 4E-F).

**Network topology regulated spike transmission and network synchronization through input convergence**

Can these results illuminate the mechanistic underpinnings of robustness and lability of breathing? In experiments, we observed variability in the latency to synchronize of the preBötC network towards full bursts, and when the same neurons were stimulated across multiple trials, there was an occasional failure to induce inspiratory bursts (Kam et al., 2013b). Similarly, in some simulation trials, activated neurons failed to induce a network burst (Figure 3 A-C). Moreover, in simulation trials where bursts did occur, multiple bouts of spiking from stimulated neurons were required (the spiking interval for each neuron was their experimentally measured interspike interval ± jitter ≈40 ± 5 ms in Figure 3A), as spikes in initial intervals were unable to induce a network burst. With respect to the example illustrated in Figure 3, when the 3$^{rd}$ set of spikes from the stimulated neurons arrive at ~100 ms in trial R1 (black), quiescent neurons were recruited but activity failed to propagate (black box, Figure 3 A-C). However, in trial R4 (blue), spikes from the stimulated neurons induced a network burst even when they were less synchronous than trial R1 (Figure 3 B-C black boxes). Similarly, in trial R5 (purple) at ~150 ms, the stimulated neurons recruited several of their postsynaptic neurons, but the network failed to fully synchronize (purple box Figure 3 B-C). In the same trial at ~200 ms, weaker synchrony among the stimulated spikes (as compared to the prior spikes at ~150 ms) induced a network burst. Based on the properties of synchronous propagation of pulse packets (Diesmann et al., 1999) with the same set of input neurons, we expected that activity propagation should be dependent only on the initial synchrony of active neurons. In contrast, we observed failure to produce bursts even when the synchrony among the stimulated neurons was higher than in some successful intervals, revealing that for the preBötC model network, local connectivity of spiking neurons and their postsynaptic partners interacts with the instantaneous network dynamics to shape the flow of activity.

To better understand the success and failure of activity fluctuation that ultimately triggers a network burst, we computed a statistical topological parameter, $S$, associated with the efferent synapses of each individual neuron (Figure S4) representing either the sum of all efferent synaptic weights with which a neuron connects to its postsynaptic neighbors ($S_{weight}$), or its total number of efferent projections ($S_{synout}$) (Figure S4A-D). The motivation to compute $S$ was to illuminate the mechanisms underlying the spread of network synchronization by coincidence detection (Figure 4D) that is dependent on input convergence (Kumar et al., 2010). We reasoned that if a set of recruited neurons had more efferent synapses, there would be a higher probability for their inputs to converge onto their postsynaptic neurons, especially in a sparsely connected network like the preBötC. Thus, we hypothesized that a high $S$ of recruited neurons was critical to induce a synchronization trajectory of the network subsequent to the initial activation of stimulated neurons.

To test this hypothesis, we plotted $S$, i.e., $\Sigma S_{weight}$ and $\Sigma S_{synout}$ (where the sum is over the active neurons), and as a measure of synchrony (Diesmann et al., 1999), $\sigma(I)$, the standard deviation of spike times of active neurons in given intervals (as defined below). In each simulation, the first interval started with the first spike within the set of stimulated neurons and ended with their last spike or that of any recruited neurons. For example, in Figure 5A the first interval, $I_1$, spans the first set of stimulated spikes of 7 randomly selected neurons (indicated by horizontal dashed lines), as their stimulation did not recruit any other neurons. The next interval, $I_2$, begins with a second stimulation of the same 7 neurons (but in different order) that recruited 3 postsynaptic neurons, and ends with the last spike of one of these recruited neurons. Still, the collective activity of stimulated and recruited neurons at the end of $I_2$ was insufficient to synchronize the network, as it is at the end of $I_3$. The dynamics change in $I_4$, where a percolation of activity appears that rapidly synchronized the network; as such, this was the last interval for this trial. For trials where the population did not synchronize, the last interval was capped at 7. Next, we plotted $\Sigma S$ vs $\sigma(I_n)$; n=1-7 to determine if there was a threshold $S$ for the network to synchronize, i.e., a minimum level of network activity where the emergent spiking induced by the stimulated neurons was sufficient to synchronize the network to burst. In Figures 5A and B, synchronization to bursting occurred when the summed population activity exceeded 10 spikes in 5 ms bins, i.e., ≥2 kHz/5 ms bin. To determine this threshold activity level more precisely, we computed the summed population activity in 1 ms non-overlapping bins. Next, we divided the last interval (where synchronization and bursting occur, e.g., $I_4$ in Figure 5A) into two subintervals: i) the *preburst interval* that started with the first stimulated spike and encompassed initial recruitment of quiescent neurons up to the 1 ms bin with ≥2 spikes and no decrease in the activity thereafter, e.g., $I_{4A}$ in Figure 5A. We reasoned that the preburst intervals contained a threshold number of neurons, which varied from trial to trial (see below) that committed the network towards synchronization. Notably, for all but one trial, i.e., R5, the first instance of instantaneous network activity with ≥2 spikes/ms, in the preburst interval, was always followed by a progressive increase in the activity leading to a burst. ii) the *burst interval* where the network fully synchronized to burst within 3-8 ms following the preburst interval, e.g., $I_{4B}$ in Figure 5A. A plot of $\Sigma S$ vs $\sigma(I)$ revealed that the likelihood of induction of network synchronization was critically dependent on both $\Sigma S$ and $\sigma(I)$. Specifically, when $\sigma(I)$, a measure of the synchrony among the stimulated (and recruited, if any) neurons, was high but their $\Sigma S$ ($\Sigma S_{weight}$ and $\Sigma S_{synout}$) was low, the network did not synchronize (Figures 5B-C). Furthermore, across different trials, the preburst intervals (that mark the induction point of bursting) with lower $\sigma(I)$, i.e., higher synchrony, had lower $\Sigma S$, and *vice versa*, revealing the synergistic interactions between $S$ (that determines input convergence) and $\sigma(I)$ (that determines input coincidence) that propels network towards a synchronizing trajectory. For successful trials, the network trajectory converged at the minimum $\sigma(I)$ and maximum $\Sigma S$ in the burst interval, which corresponds to the synchronized state of the network. These results revealed attractor dynamics in the preBötC model network (in the $S$-$\sigma$ parameter space) that resembles the dynamics of synchrony propagation in a feedforward network (Diesmann et al., 1999).

Strikingly, the dependence of network synchronization on $S$ and $\sigma$ was similar for both $S_{weight}$ and $S_{synout}$. Since the $S_{synout}$-$\sigma$ parameter space (Figure 5 C) captured the salient network attractor dynamics, we concluded that for the preBötC model on ER graphs with lognormal weights, successful bursting was dependent on the number and strength of efferent projections of initially active neurons that determines via input convergence and coincident detection the propagation of activity. As discussed above (Figure 4E, F), a lognormal distribution of synaptic weights makes every neuron a potentially good coincidence detector. However, the response reliability of any given neuron depends on the convergence of inputs, which is, in turn, critically dependent on $S_{synout}$ of spiking presynaptic neurons. As coincidence detection was only possible when there was sufficient convergence of presynaptic inputs, the trajectory of $S_{weight}$ (a determinant for coincidence detection) followed the trajectory of input convergence determined by $S_{synout}$ (*cf.* Figure 5 B, C).

We further explored the dependence of network synchronization on $S$ and $\sigma$ by using as a case study the failure of a network to burst even when the stimulated neurons resulted in relatively high synchronous activity as compared to those that induced bursts. As noted above, (Figure 3B, C),

R5 recruited neurons at ~150 ms in turn increased synchrony (decreased σ; corresponding to the fourth circle from the bottom in R5 Figure 5 D-E), however they could not sustain and amplify the activity to burst. $S$ (especially $S_{\text{syunout}}$) of these neurons was below the mean $S$ of the network (Figures 5D, E). We consider these neurons to be bad senders, insofar as their failure to induce a burst was due to insufficient input convergence and coincidence detection onto their downstream neurons. Notably, when sufficiently good senders i.e., neurons with greater than the mean $S$ in the network, were recruited in the penultimate intervals, synchronous activation of only 16-20 neurons (in addition to the 7 stimulated) was sufficient to propel the network on a synchronizing trajectory (Figure 5 D-E).

Occasionally good senders recruited in certain intervals, e.g., R2 ~150 ms and R1 ~100 ms (Figure 5 D, E) failed to sustain/amplify activity further due to lack of sufficient synchrony, i.e., high σ in these intervals. Concordantly, a combined measure of synchrony and $S$ in a single parameter, i.e., $S/\sigma$, fully recapitulates the dependence of network bursting on $S$ and σ (Figure 5 F, G). Specifically, $S/\sigma$ was low for intervals where neurons with high $S$ were active but did not sufficiently amplify network activity towards bursting. For the intervals that did induce bursting, $S/\sigma$ was always higher than in unsuccessful intervals (Figure 5 F, G). Given the critical dependence of network synchronization on these two parameters, we sought to determine the range of $S$-σ parameter space (as presented in Figures 5 B, C) that favors network synchronization. Thus, we performed additional simulations with stimulation of up to 10 neurons with a range of their spike time jitter that resulted in a wide range of $S$ and σ, then analyzed the network trajectories corresponding to these parametric ranges (Figure 5 H, I). This mapped the $S$-σ parameter space of this network close to the bifurcation points, where the network trajectories commit to synchronized bursting. Notably, one such trial revealed that a minute shift in the $S$-σ parameter space, resulting from recruitment of only one additional neuron and a 10% decrease in σ ($\Delta\sigma = -0.5$ ms) between two successive intervals, marked one of these bifurcation points (Figure 5 H, I, arrow heads) which further illuminated the attractor dynamics of this network (Diesmann et al., 1999; Ermentrout, 1998; Kumar et al., 2010; Vogels et al., 2005).

**ER network with lognormal weights reliably reproduced preBötC bursts and burstlets**

Having tested the robustness of this model to reproduce preBötC synchronization dynamics, we further explored its utility as a generalized model of preBötC rhythmogenesis. Does the model reproduce network dynamics under simulations of various experimental conditions that produce preBötC burstlets and bursts (as presented in Figure 1B)? Notably, preBötC Type I inspiratory neurons, which are putatively rhythmogenic (Ashhad and Feldman, 2020; Gray et al., 1999), start firing at very low frequencies (~0.5-1 Hz) in the late interburst interval, referred to as the preinspiratory (preI) period. Their activity leads to progressive network synchronization in the preI period that culminates in an inspiratory burst. We tested the ability of the ER network to synchronize by initializing the neurons in the network to fire randomly at Poisson-distributed frequencies (mean range: 0.5 to 2.0 Hz), thus "seeding" the network with low levels of activity, as observed preceding each burst in experiments (Gray et al., 1999). After this initialization of neuronal activity, we computed network firing rate in 5 ms bins as a measure of instantaneous synchrony in the network (Riehle et al., 1997). We chose this temporal window to capture correlated firing, i.e., synchrony, among neurons as this window (of coincidence) was too short for any neuron to spike more than once (their refractory period was set to 3 ms in all the models, a reasonable boundary condition based on experimental data, Table S1). Next, we computed average network activity from the firing rate computed in 5 ms bins, as a two-pass moving average, with a 25 ms moving window and 5 ms steps size.

When all 1000 neurons were isolated, i.e., their synaptic weights were set to zero, and initialized to fire at Poisson-distributed frequency of 0.5 Hz, network activity fluctuated around 488 ± 96 Hz ((mean ± SD); Figure 6A (black trace)). This was close to the expected network firing rate of 500 Hz, i.e., 0.5 Hz/neuron x 1000 neurons. When these neurons were then connected with uniform synaptic weight, q.v., Figure 4, the network activity, 504 ± 115 Hz, was not significantly different from that with silent synapses ($p=0.16$; Kolmogorov-Smirnov (KS) test; Figure 6A, red traces). Changing the synaptic weights to a lognormal distribution resulted in two typical network behaviors. In one typical behavior, network activity significantly increased to a firing rate to 594 ± 181, significantly higher than both the network with silent synapses ($p=5.1\times10^{-7}$, KS test) and the network with uniform weights ($p=4.4\times10^{-4}$, KS test; Figure 6A blue trace). More importantly, there were fluctuations of network activity that crossed mean activity with silenced synapses + 3SD, (upper dashed line Figure 6) that were qualitatively similar in shape to experimentally observed burstlets. In a second typical behavior, this network fully synchronized to produce a network burst (Figure 6A *cf.* blue and green traces). This ability of the network to self-organize to produce burstlets and bursts from an initial condition of low-frequency, unsynchronized, active neurons, was also reflected in a leftward shift in the interspike interval (ISI) histogram of the network with lognormal weights as compared the one with silenced synapses (Figure 6B). The ISI for the burstlet only trial (corresponding to the blue trace in Figure 6A) was 1.6±1.6 ms (mean ± SD), whose distribution was significantly different from the distribution of ISIs for the case of silent synapses at 2.0 ± 1.9 ms (Figure 6B, *cf.* blue and black traces, $p=0.002$, KS test). Thus, even when the network did not completely synchronize to burst, there was significant reduction in the mean ISI with lognormal synaptic weights, representing an emergent network dynamic. Notably, ER networks with uniform weights did not synchronize (as in Figure 6A red trace), consequently their ISI histogram was not significantly different from that with silenced synapses (with ISI at 1.93 ±1.93 ms (mean ± SD; Figure 6B, red trace; $p=0.2$ KS test)). The ISI distribution with uniform synaptic weights was also significantly different from the network with the lognormal weights ($p=0.02$, KS test). Initialization of the neuronal firing rate at a higher Poisson rate (2 Hz) in ER networks with lognormal weights, seeding activity in only 20%, i.e., 200 out of total 1000 neurons, was sufficient to induce synchronization leading to a burst (Figure 6C); these networks did not burst when the random firing was initialized in 19%, i.e., 190 of 1000, of neurons. As the mean frequency of these neurons was decreased to 0.5 Hz, the network required activity in all neurons to generate a full burst (Figure 6C) but did not generate bursts when initializing 90% of neurons. We systematically analyzed the burst probability of ER networks with lognormal weights at various Poisson-distributed frequencies of neuronal firing, ranging from 0.25 to 2.00 Hz, for 10 different realizations of these networks over 10 different trials. Networks with lognormal synaptic weights generated bursts at lower frequencies of initial neuronal firing (with Poisson-distributed mean 0.50-0.75 Hz) compared to those with uniform weight distribution (with Poisson-distributed mean 0.75-1.00 Hz; Figure 6 D). Furthermore, the range of latencies to burst for the minimum frequency at which the networks, with either uniform or lognormal synaptic weights, synchronized overlapped (lognormal weights: range=485 ms-1096 ms at 0.5 Hz, uniform weights: range=489 ms-1376 ms at 0.75 Hz, $p=0.75$ Wilcoxon rank sum test; Figure 6E). As the fraction of randomly firing neurons was decreased to 60%, i.e., 600/1000 neurons, the frequency of neuronal firing required to reliably induce network bursts increased by 0.25 Hz for ER networks with lognormal weights (from 0.75 Hz for 1000 neurons to 1.00 Hz for 600 neurons). In contrast, for networks with uniform weights this increase was higher, at 0.5 Hz (from 1.0 Hz for 1000 neurons to 1.5 Hz for 600 neurons); Figure 6F, G). Thus, networks with lognormal weight distributions appear less susceptible to disruptive perturbations that decrease neuronal activity, i.e., they are more robust.

We further explored the response of ER networks to changes in synaptic weight distribution on the response to external perturbations, such as changes in the neuronal firing due to changes in [$K^+$] in the bath solution (Kam et al., 2013a). In a network with a lognormal synaptic weight distribution, a decrease in the fraction of randomly firing neurons resulted in an increase in the latency to burst with an increased number of intervening burstlet-like events, i.e., partial synchronization and subsequent desynchronization of network activity (Figure 6 H *cf.* black trace (1000 neurons) with green, yellow and brown traces with 800, 700 and 500 neurons, respectively). However, due to stochastic nature of neuronal firing (with Poisson-distributed mean at 1.0 Hz), the latency to burst was variable and inconsistent, e.g., Figure 6H, for the network with 600 neurons, latency to burst was 70 ms, less than networks with 700 and 800 neurons, whose latency to burst was 240 ms and 105 ms, respectively. To elaborate, even though the average latency to burst (across several trials) increased with a decrease in the fraction of randomly firing neurons (*cf.* Figures 6E, G), since the process of network synchronization is driven by the stochastic firing of neurons, this results in large variability in latency to burst when seen on trial-to-trial basis where the general trend is less apparent (Figure 6H). Nonetheless, increase in the number of burstlets with a decrease in the fraction of initially stimulated neurons recapitulates the experimental behavior of preBötC rhythmogenic microcircuit (Kam et al., 2013a). This behavior was present in networks with uniform weights as well (Figure 6I), however, the network with uniform weights failed to generate bursts when ≤70% (700 of 1000) neurons fired randomly, whereas with lognormal weights the network was more reliable and failed only when the percentage of randomly firing neurons was ≤40% (400 of 1000). Thus, preBötC network models with lognormal weights are more robust and reliable in generating bursts and burstlets from noisy low frequency neuronal firing. Furthermore, the networks follow similar trajectories in the *S-σ* parameter space, as seen before in Figure 5, before they bifurcate towards full synchronization and burst generation (Figures 6J-M). During successive burstlets, the $S$-σ trajectories oscillate at the lower boundary of $S$-σ parameter space as the network alternates between partial synchronization and desynchronization. Furthermore, when as the fraction of randomly firing neurons in the network decreased, the threshold $S_{weight}$ and $S_{synout}$, for the network to fully synchronize, increased; even when the σ was similar to or smaller than the condition with 100% randomly firing neurons (Figures 6J-K; *cf.* red and yellow traces, with 50% and 70% randomly firing neurons, respectively, with the black trace for 100% randomly firing neurons). This revealed the modulation of the activity bifurcation trajectories by the state of excitability in the network(Hahn et al., 2019; Kumar et al., 2010). Taken together, these results reveal the attractor dynamics of network synchronization where low frequency fluctuating network activity exhibits a tipping point defined in the $S$-σ parameter space for the global network synchronization.

**Determining the key parameters driving synchronization**

Global network synchronization cannot be predicted solely from the number of stimulated neurons; the process clearly depends on which neurons are initially stimulated. Since neurons are distinguished only by their connections within the network, we must look for network subgraphs that, once stimulated, lead to global synchronization in order to better understand this variation of outcomes. We have already introduced a single network property - $S$ - to quantify the role of network structure on synchronization dynamics. Here we more broadly assess the importance of related network properties using machine learning as a probe. To do this, we computed a number of quantities reflecting different activated network motifs, and then asked how well a machine learning algorithm (MLA) can predict the eventual network synchronization, based solely on input data corresponding to one (or more) of these activated network quantities. We surmised that those active network quantities producing the largest increase in the reliability of the synchronization prediction by the MLA would be those that drive synchronization.

To introduce more general network quantities related to $S$, we first denote the initial state of the network by the vector $X$, where its $i^{th}$ component is one if the $i^{th}$ neuron was initially stimulated and zero otherwise; the number of these neurons is $\sum_i X_i$. We define the matrix of synaptic weights $W$, where $W_{ij}$ is the strength of the synaptic connection from neuron $j$ to neuron $i$. These quenched random matrices describe a single instance of the network. They are generally sparse and have positive nonzero entries selected from a lognormal distribution (*q.v.*, Figure S4). The average connectivity strength of a given realization of a network is $\frac{1}{N^2}\sum_{ij} W_{ij}$, where N=1000.

With these definitions, matrix multiplication generates a new vector *Y = W\*X,* which gives the list of neurons that receive spikes from the initially stimulated ones, weighted by the appropriate synaptic strength. This vector essentially encodes the effect of the initially stimulated neurons on the rest of the network, and is a function of *S*. Since neuronal stimulation and synchronization are fundamentally nonlinear processes, a vector constructed from various powers of the components of Y, which we will call *Y$^n$* for n>0, may be a better predictor than linear *S* of the synchronizing effect of the stimulated neurons, *X*. Including such nonlinearities is one route to generalizing *S*.

One may also consider higher-order interactions, including, for example, the neurons on which Y synapse: *W\*Y=W\*W\* X.* In general, one may consider n$^{th}$-degree neighbors using n factors of **W**: $W * W * ... W * X$. Combining both the nonlinearity of the neuronal response and the role of multiple-degree neighbors in the network, we can define a general class of *S* variables:

$$\boldsymbol{S}^k_{i_1 i_2 i_3 \ldots i_k} = [\boldsymbol{W} * (\boldsymbol{W} * (\boldsymbol{W} * \boldsymbol{Y}^{i_1})^{i_2} \ldots)]^{i_k}$$

These are all generalization of *S* discussed above, which, in the generalized notation we refer to as $S^1_1$.

To assess and compare the predictive power of various generalized *S* quantities, we used CatBoost MLA (Prokhorenkova et al., 2017). We trained the model to predict whether a burst will occur based on solely on the value of a particular subset of generalized senderness quantities $S^k_{i_1 i_2 \ldots i_k}$. The networks used for training and subsequent studies of the predictive power of the MLA were constructed so that their synchronization probability was close to 50% (due to the probabilistic nature of network synchronization, 53% of the networks in the training and test sets synchronized) to make the prediction task as difficult as possible. We focus on how the MLA prediction accuracy depends on the subset of $S^k_{i_1 i_2 \ldots i_k}$ chosen. The subset that provides the greatest increase in the predictive success of the algorithm indicates that the active network motifs identified by that subset of $S^k_{i_1 i_2 i_3 \ldots i_k}$ are the strongest determinants of network synchronization.

The results of this ML study are reported fully in Table S2. We note that, while all types of senderness (*S*) input improves predictive success, nonlinear generalizations of senderness individually provided the largest increase in predictive success. For instance, first-order senderness $S_1^1$ increased predictive success to 57%. Including next nearest neighbor connections with no nonlinearities did not change the predictive success. The success rate without nonlinearities never exceeded 59% in all the cases we studied. On the other hand, the simplest nonlinear generalization of the first order senderness (excluding next-nearest neighbors) $S_2^1$ improved predictive success to 66%.

Including a combination of next-nearest neighbor interactions and nonlinearity provides the highest predictive success amongst the set of generalized senderness quantities we studied individually (N=24). The most predictive was $S_{32}^2$,, i.e., the nonlinearly amplified second-order *S*, which resulted in a 72.7 ± 1% success rate. Combinations of these generalized senderness metrics did not result in significantly higher levels of predictive success. In fact, taking all twenty-four of the quantities studied in combination improved the success rate only to 73.9 ± 0.6% , barely in excess of the single most predictive measure in isolation.

The dramatically enhanced efficacy of nonlinear generalizations of senderness enables us to understand the key role of the lognormal synaptic weight distribution in network synchronization. The linear first-order senderness $S_1^1$ is proportional simply to the average synaptic weight of the activated neurons, but all of the non-linear senderness generalizations are sensitive to the infrequent but exceptionally large synaptic weights associated with the fat tail of the lognormal distribution.

**DISCUSSION**

Our analysis of the synchronization dynamics of the preBötC network model demonstrates a number of features that underlie the emergent, global synchronization of the preBötC microcircuit, either in response to the stimulation of a randomly chosen small number (≤10) of neurons, or in response to the stochastic neuron activity. We suspect that these features are widespread, applying not only to the preBötC, but to the more general phenomenon of rhythmogenesis, population synchronization and attractor dynamics of neuronal networks.

Using a simplified neuron model, only a small set of network architectures generated dynamics *quantitatively* consistent with experimental data, specifically in terms of the increasing probability of the network to synchronize as a function of the number of initially stimulated neurons and in the monotonically decreasing lag time between stimulation onset and synchronization as the number of stimulated neurons increased. Amongst the physiologically-plausible putative network architectures that we considered, we found that hierarchical networks and localized networks (those in which the presence of a synaptic connection between neurons depends on their position in space) are incapable of generating dynamics consistent with the experimental data. In contrast, dynamics on ER networks with lognormally distributed synaptic strengths, in which the probability of directed graph edges is fixed independent of interneuronal distance can produce physiologically reasonable dynamics that resemble experimental observations. Moreover, when we perturb the ER network topology by increasing the number of 3-simplices, as has been suggested for other neuronal networks (Gal et al., 2019; Perin et al., 2011), these networks became too sensitive to exogenous stimulation as compared to the experimental data. In studying the dynamical trajectories of the networks towards synchronization, we found a predictive two-dimensional parameter space spanned by spike-timing synchronization (as measured by its variance) and the total efferent synaptic connections/weights, *S,* of spiking neurons (Figures 5-6).

Notably, these relatively simple models of excitatory networks were able to capture the experimentally observed preBötC dynamics is also due to the fact that the core preBötC rhythmogenic microcircuit is excitatory (Ashhad and Feldman, 2020; Cui et al., 2016; Del Negro et al., 2018; Feldman and Kam, 2015; Kam et al., 2013b) and under the experimental conditions modelled here, with the caged MNI-Glutamate and 9 mM [$K^+$] in the bath solution, $GABA_A$ inhibition is blocked (Ellis-Davies, 2018) and network excitability is high. Furthermore, even though the probability to generate an I-burst in these experiments was ≳80% (Kam et al., 2013b), in vivo the network output is highly robust due to the fact that activity initiating from several loci within the preBötC, during preI, can converge to induce a global synchronization (Ashhad and Feldman, 2020; Kam et al., 2013b).

We emphasize that network architecture alone is not sufficient to produce the physiologically relevant synchronization dynamics; the distribution of synaptic weights is critical. Networks with a uniform distribution of synaptic weights exhibit, in fact, poor synaptic efficacy towards coincident inputs, i.e., they require more simultaneously stimulated neurons to globally synchronize than required experimentally. By including a lognormal distribution of synaptic weights in our model, we produced networks whose sensitivity to exogenous stimulation matched that observed experimentally.

To further test this point, we used machine learning to probe which activated network properties are the best predictors of global synchronization. We explored generalizations of efferent connections, senderness, by considering next-nearest neighbor (and higher order) interactions and incorporating nonlinear neuronal responses. All linear variants of *S*, which are insensitive to the nonlinear, all or none, spike generation process in neurons, performed poorly as predictors of synchronization, while nonlinear variants were significantly better predictors. In particular, a generalized senderness parameter that includes both next-nearest neighbor interactions and nonlinearities was the best predictor. This further supports the centrality of large synaptic weights in the fat tail of the lognormal distribution, which represent only a small fraction of all connections, in determining the collective dynamics of the system.

The networks identified as generating a physiologically reasonable response to external stimulation also produced physiologically consistent stochastic dynamics in response to noisy spiking of their constituent neurons. Importantly, the activity emerging from stochastic neuronal firing could reproduce burstlets, i.e., low levels of network activity above noise, as seen experimentally (Kam et al., 2013a) that we hypothesize are kernel of breathing rhythmogenesis (Del Negro et al., 2018; Feldman and Kam, 2015; Kam et al., 2013a). In our models, burstlets are the result of incomplete synchronization and subsequent desynchronization of the network and suggests that for burstlet termination, inhibition, feedback or otherwise, is not obligatory.

Another critical feature underlying oscillatory dynamics in the preBötC of our successful models is the lognormal synaptic weight distribution. Specifically, neurons connected through lognormal weights exhibited increased synaptic efficacy as well as differentiators of synaptic inputs from their neighbors (Figure 4). While coincidence detection promotes synchrony, the differentiator property allows neurons to dampen the network activity by suboptimal firing when the impinging synaptic inputs are weaker than the network average (Figures 4E, F). The reduced firing of these neurons then leads to desynchronization of partially synchronized subthreshold network activity. Thus, by providing a broader fluctuation range for network activity, which leads to better synchronization probability, lognormally distributed synaptic weights impart a dynamical property to local network motifs, controlling both the yin and yang of oscillatory dynamics(Gore and van Oudenaarden, 2009). This reveals a synaptic mechanism underlying robustness of preBötC output even when in each cycle the inspiratory bursts emerge from low levels of fluctuating activity with variable lags constituting the preI duration.

The models with uniform synaptic weights could also partially synchronize and then desynchronize as the spiking neurons, differing in their efferent connectivity, constituted dynamic motifs (Womelsdorf et al., 2014) with varying degree of input convergence. However, if instead synaptic weights were lognormally distributed, this type of oscillatory dynamics was achieved with lower firing rates of individual neurons. We conclude that dynamically emerging circuit motifs of active neurons interacting via lognormally weighted synapses can generate oscillatory dynamics that parallel those observed experimentally. The ability of the network to desynchronize to baseline activity levels suggests that the preBötC burstlet may have different termination mechanisms compared to those of the bursts, which may rely on other properties (Del Negro et al., 2018; Baertsch et al., 2018; Del Negro et al., 2005; Janczewski et al., 2013; Kottick and Del Negro, 2015; Krey et al., 2010; Pace et al., 2007b; Rubin et al., 2009).

Taken together, our results show that the observed fluctuations of the isolated system and its response to small perturbations, i.e., stimulation of <1% of the network, provide important constraints on the neuronal architecture of the network. Our numerical studies also suggest that one may further test the model by looking for activated network motifs involving only first and second nearest neighbors as a necessary prelude to global network synchronization.

One may ask why an apparently complicated network topology could be an essential element for rhythmogenesis when much simpler models, e.g., those involving pacemakers and/or limit cycles, can fit much of the experimental data. Setting aside the issue of synchronization, both classes of models appear robust. We posit that our network-based burstlet hypothesis is significantly more flexible, a critical property of breathing in mammals to provide the rapid response necessary to cope with real world perturbations, e.g., rapidly increased metabolism during exercise such as when fleeing a predator or essential reflexes such as sneezing and coughing. Concordantly, our study reveals that a fat tailed synaptic weight distribution renders the preBötC rhythmogenic network stable yet more labile in response to external inputs. Similar robustness of activity propagation and oscillatory dynamics are present in other brain networks that exhibit lognormally distributed synaptic strengths (Buzsaki and Mizuseki, 2014; Ikegaya et al., 2013; Omura et al., 2015; Song et al., 2005; Uzan et al., 2018). Given that the mammalian breathing pattern needs to be spontaneously tuned with a wide range of behaviors such as vocalization, speech, swallowing, locomotion and emotion, e.g., laughing, crying, sighing, we reason that a network-based mechanism for inspiratory rhythms that relies on neuronal synchrony (Ashhad and Feldman, 2020; Srivastava et al., 2017) and exhibits attractor dynamics (Diesmann et al., 1999; Kam et al., 2013a; Kam et al., 2013b; Uzan et al., 2018) best captures the physiological boundary conditions as well as the experimental data on preBötC rhythmogenesis.

In conclusion, we demonstrate the plausibility of our hypothesis that a robust and reliable transformation of preBötC network activity from random to synchronized underlies rhythmicity by burstlet and burst generation. We suggest that this mechanism of network assembly may underlie functions of microcircuits throughout the mammalian nervous system.

**ACKNOWLEDGEMENTS**
This work was supported by NIH-NHLBI grant 1R35HLI35779 to J.L.F, Bhaumik Institute Fellowship, Holmes-Peccei Fellowship and Dissertation Year Fellowship to V.S., NSF-DMR1709785 to A.J.L. and V.S. We thank Profs. D.V. Buonomano, Larry Abbott and Divyansh Mittal for their critical reading and insightful comments on earlier versions of this manuscript.

**AUTHOR CONTRIBUTIONS**

J.L.F and A.J.L conceptualized the project, J.L.F, A.J.L, V.S and S.A designed the study, V.S and S.A performed simulations and analyzed data. All authors critically evaluated data, co-wrote paper and approved the final version of the manuscript.

**DECLARATION OF INTERESTS**

The authors declare no competing interests.

# METHODS

## LEAD CONTACT AND MATERIALS AVAILABILITY

Further information and requests for resources and reagents should be directed to and will be fulfilled by the Lead Contact, Jack L. Feldman (feldman@g.ucla.edu)

## METHOD DETAILS

### Model

### Neuronal dynamics

We modeled neurons as compact, i.e., point, leaky integrate-and-fire (LIF) neurons. The change of somatic potential $V_i$ of the $i$-th neuron with time $t$ is controlled by(Gerstner, 1995):

$$\tau_m \frac{dV_i}{dt} = V_{rest} - V_i(t) + R \sum I_{i,j}(t) \quad (1)$$

where $V_{rest}$ is the resting potential with no external input, $\tau_m$ is the membrane time constant, $R$ is the input resistance, $I_{i,j}$ is the input current from the $j^{th}$ to $i^{th}$ neuron, the sum is over all input neurons that synapse to $i^{th}$ neuron. $V_{rest}$ was set at either at –60 mV or –65 mV depending on the simulated experimental conditions. The current is defined as(Brunel and Sergi, 1998):

$$I_{i,j}(t) = \begin{cases} \frac{W_{ij}}{R}(t - t_j - \Delta t_{ij}) \exp\left(-\frac{t - t_j - \Delta t_{ij}}{\tau_s}\right), & t \geq t_i + \Delta t_{ij} \\ 0, & t < t_j + \Delta t_{ij} \end{cases}$$

Where $t_j$ is time of spike initiation in the $j^{th}$ neuron, the synaptic weight $W_{ij}$ controls the magnitude of the spike going from the $j^{th}$ neuron to the $i^{th}$ neuron, $\tau_s$ sets the temporal width of the spike, and $\Delta t_{ij}$ is the transmission delay time for the spike to travel from neuron $j$ to neuron $i$.

For single spikes, these equations can be solved analytically. Assuming neuron $i$ to be at its resting potential before the arrival of a spike, the solution for its potential (which is the waveform of the resulting EPSP) is

$$V(t) = \gamma_1 W_{ij} \frac{e^{-\gamma_1 t'} - e^{-\gamma_2 t'}(1 + \gamma_3 t')}{\gamma_3^2}$$

with $t' = t - t_i - \Delta t_{ij}$, $\gamma_1 = 1/\tau_m$, $\gamma_2 = 1/\tau_s$, $\gamma_3 = \gamma_2 - \gamma_1$.

While $\Delta t_{ij}$ can be taken directly from the experimental data (Rekling et al., 2000), we need to fit parameters $\tau_m$, $\tau_s$, $W_{ij}$ to obtain approximately the same EPSP profile as observed experimentally. The EPSP decay time constant (Rekling et al., 2000) is almost $\tau_m$ ($\tau_m \gg \tau_s$). We then chose $\tau_s$ to fit the EPSP rise time. The lognormal distribution for EPSP amplitudes was determined from published parameters (Rekling et al., 2000). Specifically, the unitary EPSP amplitude of inputs from putative rhythmogenic neurons was distributed as 2.8 ±1.5 mV (mean±SD). The high variance of EPSP amplitude requires a heavy tailed distribution in order to capture the entire range of EPSPs without incurring negative values, i.e., assuming a symmetric normal distribution 99.75% (mean±3SD) range of EPSPs would be -1.7 mV to 7.3 mV. Furthermore, this EPSP distribution of preBötC rhythmogenic neurons is consistent with the notion that biologically plausible, lognormal distributions of unitary EPSP amplitudes seen in several brain regions (Buzsaki and Mizuseki, 2014) are essential to explain their behavior. Thus, we modeled EPSP amplitudes by fitting the parameters of a lognormal distribution for synaptic weights $W_{ij}$ to reflect the experimental range of unitary EPSP amplitudes. In terms of the model parameters, an EPSP amplitude of 2.8 mV required a synaptic weight of $W_{ij}$=300 mV/ms.

The deterministic generation of spikes by a model neuron was controlled as follows. When the neuronal potential exceeded a threshold $V^*$ it fires, and the potential dropped instantaneously back to $V_{rest}$, with the boundary condition that the neuron cannot fire during the refractory period after a previous spike ($\tau_{refractory}$=3 ms)(Ashhad and Feldman, 2020; Ermentrout, 1998; Gerstner, 1995) ; the potential then continued to obey Eq. (1). When, however, we allowed spontaneous stochastic firing (Figure 6 only), regardless of potential, spontaneous firing may occur due to inherent neuronal excitability. This baseline firing was modeled as a Poisson process with the frequency $f_{noise}$ (0.5 to 2.0 Hz), creating background uncorrelated network activity(Gray et al., 1999). We demonstrate the dynamics of the model in a network with three neurons (Figure S1).

Finally, we simulated the effects of experimental photostimulation of preBötC neurons by "depolarizing" chosen neurons to produce a similar spiking pattern(Kam et al., 2013b). We characterized the spiking pattern by the time between activation and the first spike $\tau_{delay}$, average period of spiking of the laser stimulated neurons, $T_{laser}$, the standard deviation, $\Delta T$, of the spiking period distribution, and the number of spikes produced $n_{spikes}$, all based on experimental measurements (Kam et al., 2013b). The parameters used for the model are listed Table S1.

**Simulations for stimulation of 1-10 preBötC neurons leading to network bursts**

Stimulation (by holographic photolysis of caged glutamate (Kam et al., 2013b)) of 4 – 9 of preBötC neurons induces inspiratory bursts (I-burst) through synchronization of preBötC rhythmogenic neurons with a ≳80% success rate, with latencies ranging from 57 ms – 360 ms that are inversely related to the number of stimulated neurons (Figure 1 D-G). Specifically, the minimum number of co-stimulated neurons that induced an I-burst (henceforth referred to as the threshold number of neurons) ranged from 4 – 9 with the latency to I-burst generation ranging from 170 ms – 370 ms (255±43 ms; mean±SEM) (Figure 1H). For experiments with suprathreshold number of stimulated neurons, consisting of 1 – 3 additional neurons, the latency to induce I-burst ranged from 57 ms – 160 ms (125±23 m; mean±SEM; Figure 1I) (Kam et al., 2013b).

To model the experimental conditions in a connected population of 1000 LIF point neurons (the same order as the number of preBötC rhythmogenic neurons), we mimicked the effect of 9 mM [$K^+$] in the extracellular bathing solution on neuronal excitability by incorporating a constant depolarizing potential that put $V_{rest}$ ~ –60 mV while keeping the threshold for spikes at – 48 mV; for the purposes of the model, this was represented by a 12 mV depolarization from baseline potential, defined as 0 mV. To simulate the holographic photolysis experiments(Kam et al., 2013b), we initiated spiking in 1-9 neurons in a connected population(Rekling et al., 2000) of 1000 LIF neurons, q.v., MODEL section, at 25±3 Hz per their experimentally characterized firing behavior(Kam et al., 2013b) (Figure 1F).

**Network structure**

To complete our model definition, we need to specify the fixed network over which the neurons interact. As pointed out above, the real network topology does not appear to exhibit any anatomical and cytoarchitectural regularity, as distinguished from such structures as cortical barrels or hippocampus. Consequently, we investigated a few classes of physiologically plausible networks. Specifically, we consider four distinct ensembles of random directed graphs: (i) *Erdős–Rényi (ER)*(Erdos, 1959; Gilbert, 1959) directed graphs; (ii) directed graphs with an increased number of directed 3-simplicies (*small world*; see definition below); (iii) *localized*, and; (iv) *hierarchical* graphs.

In the ensemble of directed ER graphs, the probability for any two rhythmogenic preBötC neurons to be unidirectionally connected was the same, i.e., $p=0.065$ based on experimental data (Rekling et al., 2000). Bidirectional connections, which were not observed experimentally (Rekling et al., 2000), were not forbidden in this model, though their probability $p^2$ was negligible.

Next, we considered networks in which the number of directed 3-simplicies was higher than in *ER*, i.e., a *small world* network (Watts and Strogatz, 1998). The directed 3-simplex is a structure of three neurons connected such that neuron A synapses to both neurons B and C, and neuron B synapses to neuron C (Figure S1C). While for the *ER* network the number of directed 3-simplices is $N_s = (pN)^3$, for some neuronal networks, including neocortex, the number of directed 3-simplices is significantly higher (Gal et al., 2019; Perin et al., 2011). To create such a network, we started with the *ER* network and add 3-simplicies.

We also considered *localized* networks formed as follows: we placed ~1000 neurons in a planar, square array with neighboring neurons separated by unit distance. We then added directed edges to form a network, such that the probability for two nodes to be connected decayed exponentially with distance (measured in the usual way), i.e., the probability of a direct connection between neurons *i* and *j* was given by $p_{ij} \alpha\ exp(-d_{ij}^2/\lambda^2)$, where $d_{ij}$ is the distance between them and $\lambda$ is the mean connection length. The case $\lambda \gg 1000$ corresponds to the *ER* network. A representation of one example from this ensemble of random networks is shown in Fig. S1. The motivation for exploring this sort of network is based on the fact that synaptic connections in some compact neural microcircuits decrease with interneuronal distance(Perin et al., 2011) The model may be generalized to three dimensional networks, but we did not pursue that here.

Finally, we considered the ensemble of *(c,q)-hierarchical* networks of *N* neurons with *c* central groups of at least *q* neurons each, and one large peripheral group containing the almost all neurons of the network – $N >> cq$; *q* was chosen such that *q* simultaneous EPSPs incident on a neuron will produce a spike. Each neuron in the peripheral group synapsed on each neuron in the first central group, each neuron in the first central group synapsed on each neuron in the second central group, etc. Neurons in central groups were all to all connected, while neurons in the peripheral group were not connected to each other. Finally, each neuron from the last central group synapsed to each neuron in the peripheral group. See Figures S1C and 2E for *c=2* example. We introduced multiple central groups in order to avoid bidirectional connections between pairs of neurons, as these have not been observed experimentally (Rekling et al., 2000). These hierarchical networks were the most likely to produce a burst. Indeed, in this network the synchronous firing of *q* peripheral neurons, which make up the vast majority of the neurons, launched a burst. Meanwhile, for *any* network the synchronous firing of *q* neurons was a necessary requirement for the activation of the next neuron, but in the other ensembles of networks this condition was generally not sufficient. Therefore, there is no network we studied that was more likely to synchronize than the hierarchical one.

**Simulations and Graph Analysis**

Network models were implemented in Python with NumPy. Simulations were performed on the Jupyter Notebook platform with integration time step of 50 μs. Graph statistics (Figure 2) were analyzed in the Gephi (version 0.9.2) graph visualization platform ([www.netbeans.org](www.netbeans.org)). Data analysis was performed with custom-written software in IgorPro (version 7.08, WaveMetrics, Inc.).

**QUANTIFICATION AND STATISTICAL ANALYSIS**

Statistical analyses were performed using IgorPro (version 7.08, WaveMetrics, Inc.). For statistical significance testing normality was neither tested nor assumed and hence, non-parametric tests were used. For box-and-whisker plots, center line represents the median; box limits, upper and lower quartiles and whiskers represent 90 and 10 percentile range.

## DATA AND CODE AVAILABILITY

The codes for simulations generated in this study are available from the corresponding author upon reasonable request.

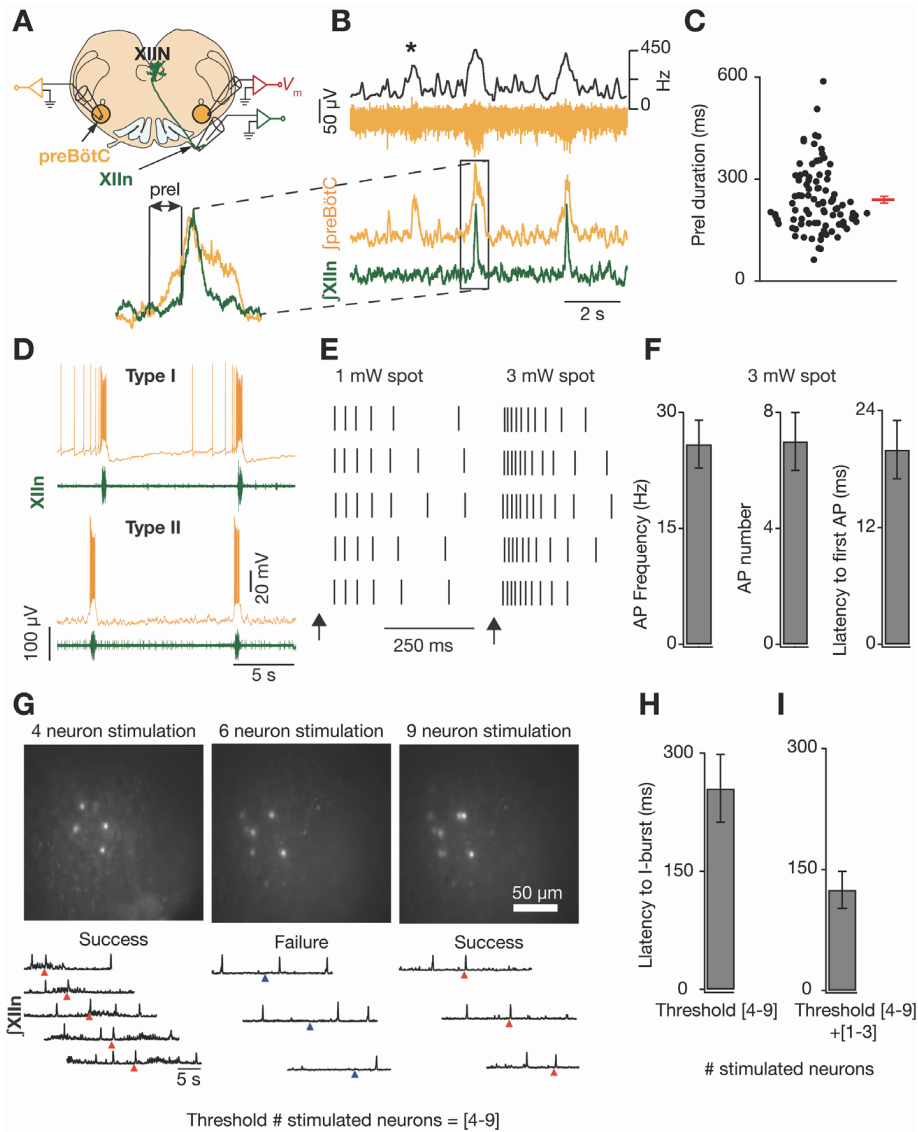

**Figure 1 preBötC population dynamics underlying inspiratory burst generation. A-C**, adapted from Ashhad & Feldman, 2020. **A**, Schematic of recording configuration from brainstem slice of neonatal mouse. preBötC: preBötzinger Complex, XIIn: hypoglossal nerve originating from XII nucleus (XIIN), $V_m$ membrane potential recording from preBötC neuron. **B**, preBötC population recording (9 mM K$^+$ bath solution) exhibiting rhythmic population bursts. Top, orange: raw recording, black: instantaneous population activity. Bottom integrated preBötC activity (∫preBötC, orange) with simultaneously recorded motor output from hypoglossal nerve (∫XIIn, green). Inset shows boxed region with early preBötC activity marked as preinspiratory (preI) duration. **C**, Bee swarm plot showing distribution of preI durations with their mean±SEM (red). **D**, adapted from Gray *et.al.*, 1999. Two types of preBötC inspiratory-modulated neurons. Type I neurons (top) start low-frequency firing during preI. Type II neurons (bottom) fire during inspiratory burst (I-burst), but not during preI. **E-H**, adapted from Kam *et. al.*, 2013b. **E**, raster plots of spikes of a recorded neuron evoked over 5 trials by uncaging MNI-Glutamate with 10 µm laser spots, as a function of laser power, beginning at arrow. **F**, Summary data for experiments in E depicting action potential (AP) frequency (left), total number of AP generated (middle) and the latency to fire first AP (right) poststimulation. n=11 neurons. **G**, examples of 3 different experiments of multiple trials with holographic photostimulation (top frames) of 4, 6 or 9 neurons with XIIn recording (bottom) in rhythmic slice preparations; photostimulation onset indicated by triangles: red triangles indicate success and blue triangles indicate failure to elicit an ectopic I-burst. **H**, Latency to induce I-burst after the onset of photostimulation in minimum number (threshold) of stimulated inspiratory-modulated neurons, ranging between 4-9, required to induce an I-burst (H). **I**, same as G but with additional 1-3 stimulated neurons. For H-I, n=4 experiments with 5-10 trials each. All data expressed in mean±SEM,

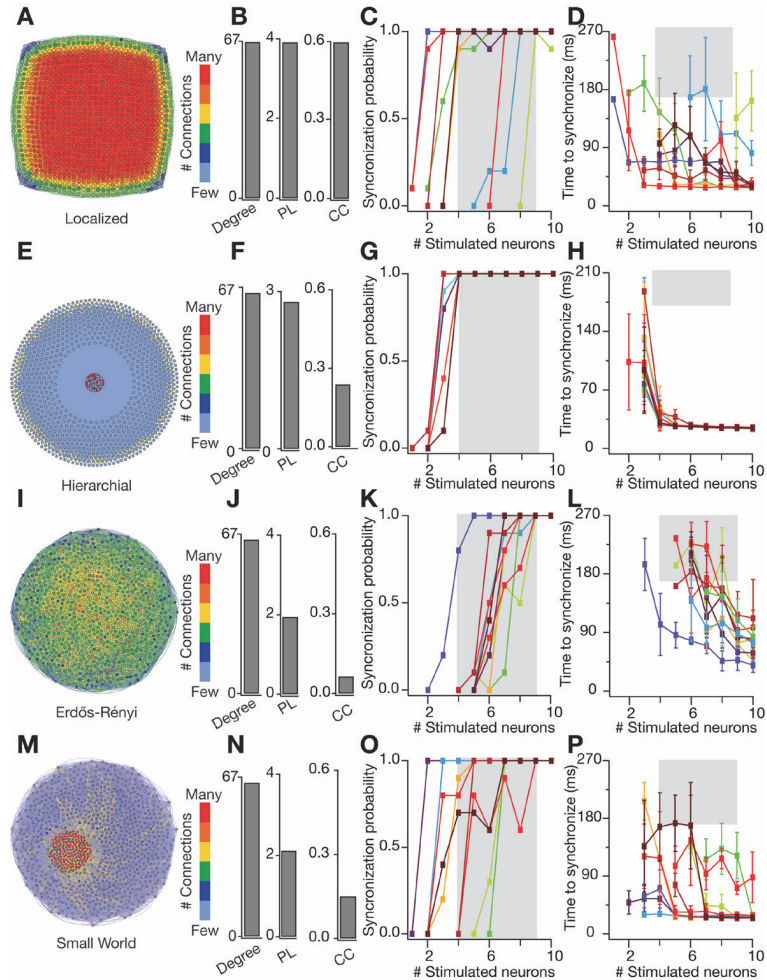

**Figure 2. Randomly connected networks of excitatory neurons replicated experimentally observed preBötC synchronization. A**, localized networks on force-based (Fruchterman-Reingold) layout where linked nodes (neurons) are *pulled* together and unrelated nodes are farther apart; most strongly linked nodes (through direct connections or common inputs) are at center and the least linked ones at periphery. Nodes (and their edges) are color coded based on number of their projections, i.e., outward synapses with warmer shades representing more connections. Higher resolution figures in Supplementary material. **B**, Degree, average path lengths (PL) and average clustering coefficients (CC) for localized networks used in this study. **C-D**, synchronization probability (C) and mean latency to synchronize (D) for localized networks when (1-10) neurons were stimulated to replicate experimental protocol described in Figure 1. Each colored trace represents a different network where synchronization probability and the mean latency to synchronize was computed over 10 trials; grey boxes span the parameter space that lie within the experimental range (170 ms-370 ms) (Kam *et.al.*,2013b) for threshold number of stimulated neurons to induce preBötC bursts. **E-H**, same as A-D, respectively, but for hierarchical 2-center networks (see METHODS); here the neurons of the two centers (as depicted in Supplementary Figure 1 are pulled together into the center of the graph but can be distinguished based on their projections. **I-L**, same as A-D, respectively, but for random networks modeled on ER graphs. **M-P**, same as A-D, respectively, but for networks with increased number of triplices (triangular directed edges as described in Figure S2) incorporated in the ER network, resulting in small world network. Note that only ER networks replicate the experimental finding of minimum number of stimulated neurons required to induce preBötC burst to be between 4-9 (K), with latency to synchronize exhibiting the widest range, closely representing the experimental results. For C, G, K and O, across 10 different realization of each network type, minimum number of neurons required to synchronize the network with ≥80% reliability = 2-9, 3-4, 4-9 and 2-7, respectively. For D, H, L, and P range of mean latency to synchronize the network with the minimum number of stimulated neurons at ≥80% reliability = 55-145 ms, 43-143 ms, 88-187 ms and 27-121 ms, respectively; error bars = standard deviations.

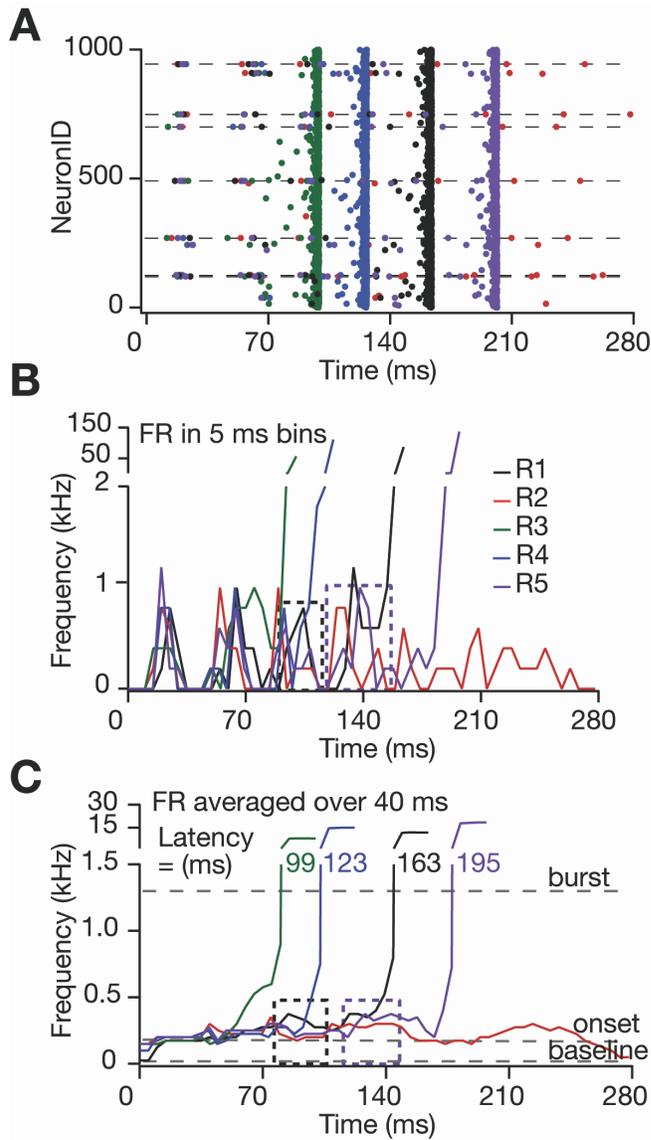

**Figure 3. ER graphs with lognormal synaptic weights reproduced the robustness of preBötC synchronization and trial to trial variability in the latency to synchronize seen in experiments**. **A**, Model output when the same set of randomly selected 7 neurons (on the dashed lines) was stimulated to fire seven spikes each at 25 Hz with 5 ms jitter, i.e., SD to model holographic uncaging of glutamate onto these neurons as in (Kam et. al., 2013b). Spike times for all neurons represented by circles color coded for 5 trials (green, blue, black, purple, red)). Y-axis represents arbitrary order of 1000 LIF preBötC neurons. In these trials, neurons #121, 126, 270, 491, 700, 749, 943 were stimulated. In 4/5 trials (green, blue, black and purple), the network synchronized, indicated by temporal alignment of spikes in all neurons, but at various latencies. In the 5th trial (red), the network did not synchronize, i.e., no vertical alignment of red dots. **B**, firing rate (FR) of stimulated neurons and their postsynaptic activated neurons in A (color coded as in A) in 5 ms bins showing waxing and waning of their activity during and poststimulation. **C**, average firing rate of network computed by averaging network activity in a moving window of 40 ms with 5 ms step increment. Dashed boxes in B and C represent unusual intervals where despite of relatively high synchronous activity, emerging from stimulated and their recruited neuronal firing, the network did not synchronize fully.

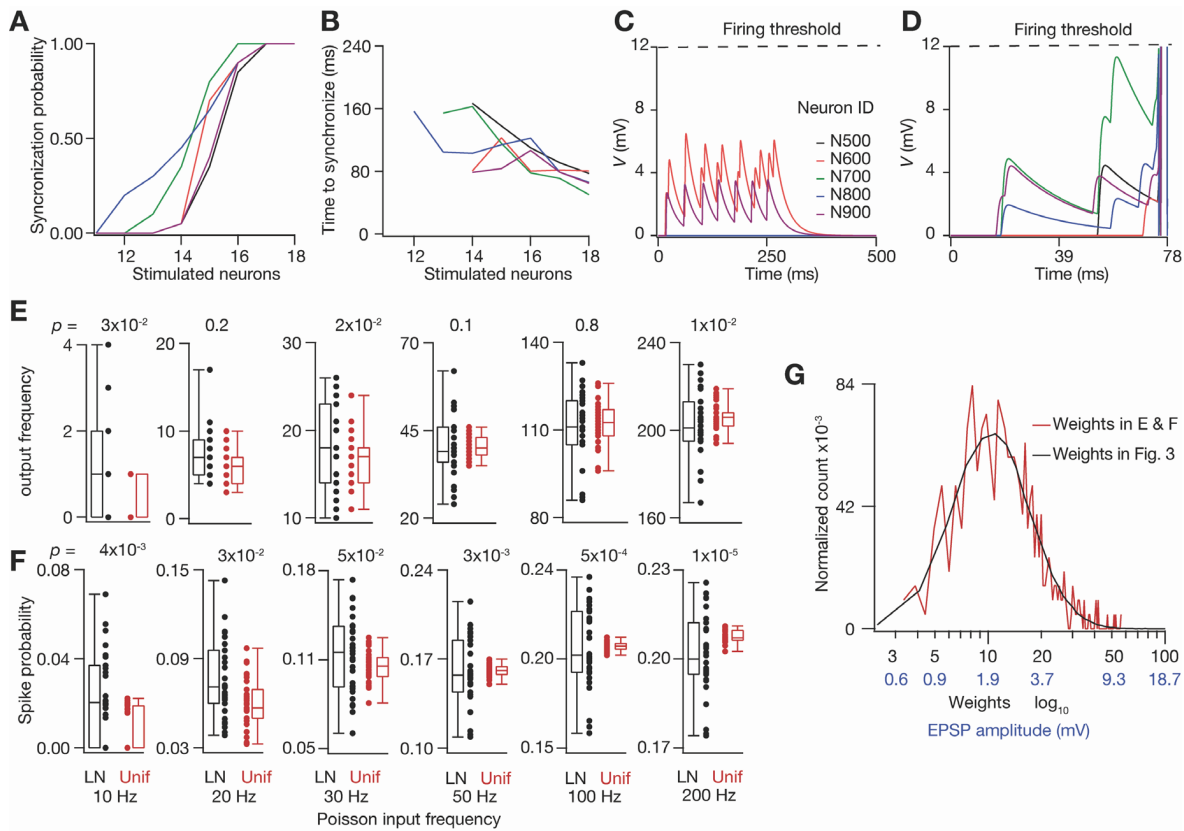

**Figure 4. Lognormally distributed synaptic weights enhanced the fidelity of spike transmission that enables network synchronization with fewer inputs. A-B**, Synchronization probability (A) and latency to synchronize (B) for 5 different ER networks with uniformly distributed synaptic weights that were equal to mean weight used in Figures 2-3. **C,** voltage of 5 randomly selected neurons from an ER network with uniform weights when a different set of randomly selected 7 neurons was stimulated like simulations in Figure 4A. **D,** same as C when the same network connectivity was incorporated with lognormal weight distribution. The network synchronized at ~78 ms and voltage traces reveal better coincidence detection in this network; vertical lines ~78 ms represent action potentials. **E-F**, average firing frequency (E) and spike probability (F) of individual neurons when 10 randomly selected synaptic inputs (out of 50) were activated at 6 different Poisson frequencies (10-200 Hz; indicated at bottom of F), with either lognormal (LN) or uniform (Unif) synaptic weight distributions; these are composite results from 3 trials each of 10 different neurons at each stimulation frequency; $p$ values for KS test. **G** histogram of LN weights used E and F (red) compared with the distribution of weights of Figure 3; corresponding EPSP amplitudes for the weights are indicated in blue.

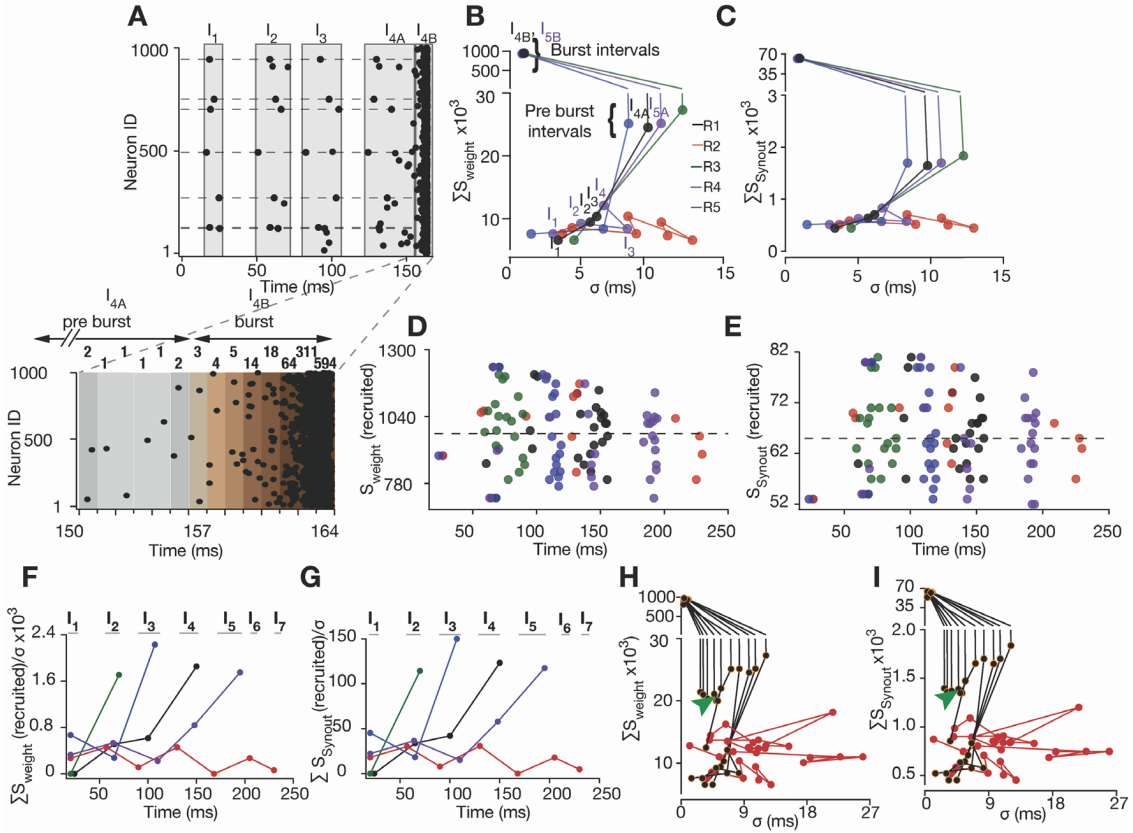

**Figure 5. Network topology regulates spike transmission fidelity and network synchronization through input convergence. A,** Illustration of interval selection for an example trial (R1 of Figure 3) of spike times after 7 randomly selected neurons were stimulated to fire up to 7 spikes at 40 ± 5 ms (mean ± SD) period. Spikes in these intervals were used to compute S and σ for plots in (B-I) as described in the text. Intervals $I_1$-$I_3$ enclose the spike times of the stimulated neurons (demarked by dashed horizontal lines) as well spikes of their postsynaptic recruited neurons. Penultimate (pre-burst) interval ($I_{4A}$ for R1) encloses the last set of spikes from stimulated and recruited neurons up to the last 1 ms bin, at which point the network activity always increased monotonically. Inset, below, shows expanded region from top, illustrating the selection of end boundary for the pre-burst interval; shaded rectangles represent 1 ms bins with darker color representing higher population firing frequency (kHz) denoted at the top of each bin. The final interval ($I_{4B}$ for R1) represents the synchronized state of the network; see text for details of interval selection for various trials. **B-C** $S_{weight}$ (B) and $S_{synout}$ (C) plotted against σ for each trial of Figure 4A, color-coded for individual trials. Each circle corresponds to the results for 1 interval in a single trial and lines connect evolution of network activity across successive intervals. Standard deviation of spike times in each interval (σ) is a measure of synchrony among the spiking neurons in that interval; intervals for two trials (R4 & R5) are indicated in color code for illustration. **D-E,** $S_{weight}$ (D) and $S_{synout}$ (E) of recruited neurons up to the pre-burst interval, i.e., the onset of network synchronization without any further external stimulation. Dashed line represents the mean S of the network. Note the failure of purple trial to synchronize in interval $I_4$ (~150 ms) even when it has high synchrony/low σ (in B and C). Here the S of recruited neurons is low, thus the network did not synchronize. **F-G,** $S_{weight}$/σ and $S_{synout}$/σ of neurons recruited by activity of stimulated neurons in each interval (indicated at the top) across the five trials. Note that for the activity in the intervals that did not induce network synchronization in the network the S/ σ was always lower than the ones that led to network synchronization. **H-I,** $S_{weight}$, (H) and $S_{synout}$ (I) vs. σ, as in (B-C) with additional simulations showing parameter space where the network activity bifurcates towards synchronization. For these simulations 7-10 neurons were stimulated to fire 7 spikes at 40-100 ms mean period with 5-8 ms jitter (SD). Red trials did not synchronize. Green arrowheads represent a trial where small change in the (S-σ) parameter space in the successive interval (with recruitment of only one additional neuron and Δσ= –0.5 ms) resulted in the bifurcation of network trajectory towards synchronization.

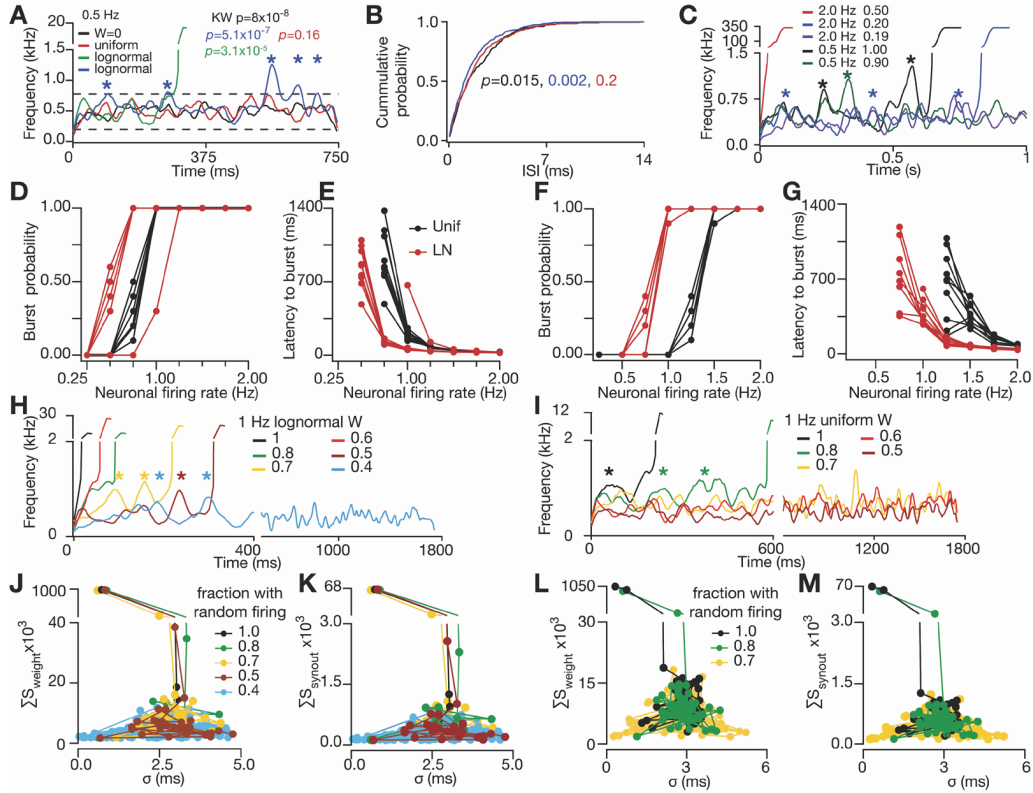

**Figure 6. ER network with lognormal weights can reliably generate preBötC bursts and burstlets. A,** average firing rate of various ER networks with different synaptic weight regimes (W: no connections (black); uniform: uniform connections (red); lognormal (LN): lognormal connections (green, blue, different trials on the same network)) when all neurons were initialized to fire at Poisson distributed frequency around 0.5 Hz. Dotted lines mark 99.75% range of firing rate for the network with synaptic weight (W)= 0. For one network with LN weight distribution (blue), note the partial synchronization and subsequent desynchronization of network activity (*) that is similar to experimentally observed burstlets (Kam *et.al,* 2013b). **B,** normalized cumulative probability distributions of interspike interval (ISI) for traces in A; for A&B, *p* value for Kruskal Wallis test (black) followed by KS test for pairwise comparisons of network activity (A) and ISI histograms (B) (color-coded to respective conditions) with W=0 condition. **C,** average firing rate of different networks when various fractions of neurons (range: 0.19-0.90; indicated to the right of firing frequency legend) were initialized to fire at Poisson distributed frequency around 2 Hz or 0.5 Hz. Note the higher amplitude of burstlets (*) at lower "noise" (0.5 Hz; green and black), as compared to the ones at 2 Hz in blue and purple). **D-E,** probability (D) and mean latency to burst (E) of 10 ER networks with uniform (black) and LN weight distribution (red) when all neurons were made to fire at various Poisson distributed frequencies (Neuronal firing rate) indicated. **F-G,** same as D-E, respectively but with only 60% (600) of the neurons were initially activated. Note in D and F that the networks with LN weights synchronized at lower frequency of neuronal firing as compared to the one with uniform weights. **H,** firing rate of an ER network with LN weight distribution when various fractions of its neurons (as indicated in color-code) were activated to fire at Poisson distributed frequency around 1 Hz. **I,** same as H but with uniform weight distribution; note that the partially synchronized network activity (burstlets, ), preceding the full network burst, are more prominent and can be generated with a lesser fraction of randomly spiking neurons in networks with LN weights (H) as compared to the ones with uniform weight (I). **J-K,** $S_{weight}$-σ (J) and $S_{synout}$-σ (K) parameter space of traces in H. **L-M,** $S_{weight}$-σ (L) and $S_{synout}$-σ (M) parameter space of traces in I. For J-M, S and σ was computed in non-overlapping 10 ms windows. J-M reveal the attractor dynamics of network synchronization. $V_{rest}$ for these simulations was fixed at –65 mV ($V^*$ =17 mV) to match low frequency firing of the type1 inspiratory neurons near resting potentials (Ashhad and Feldman, 2020; Gray et al., 1999) during pre-inspiratory period.

# Microcircuit synchronization and heavy tailed synaptic weight distribution in preBötzinger Complex contribute to generation of breathing rhythm

Authors: Valentin M. Slepukhin[a]*, Sufyan Ashhad[b]*, Jack L. Feldman[b][§], and Alex J. Levine[a,c,d][§].

[§]Corresponding Authors: feldman@g.ucla.edu, alevine@physics.ucla.edu

Supplemental Information

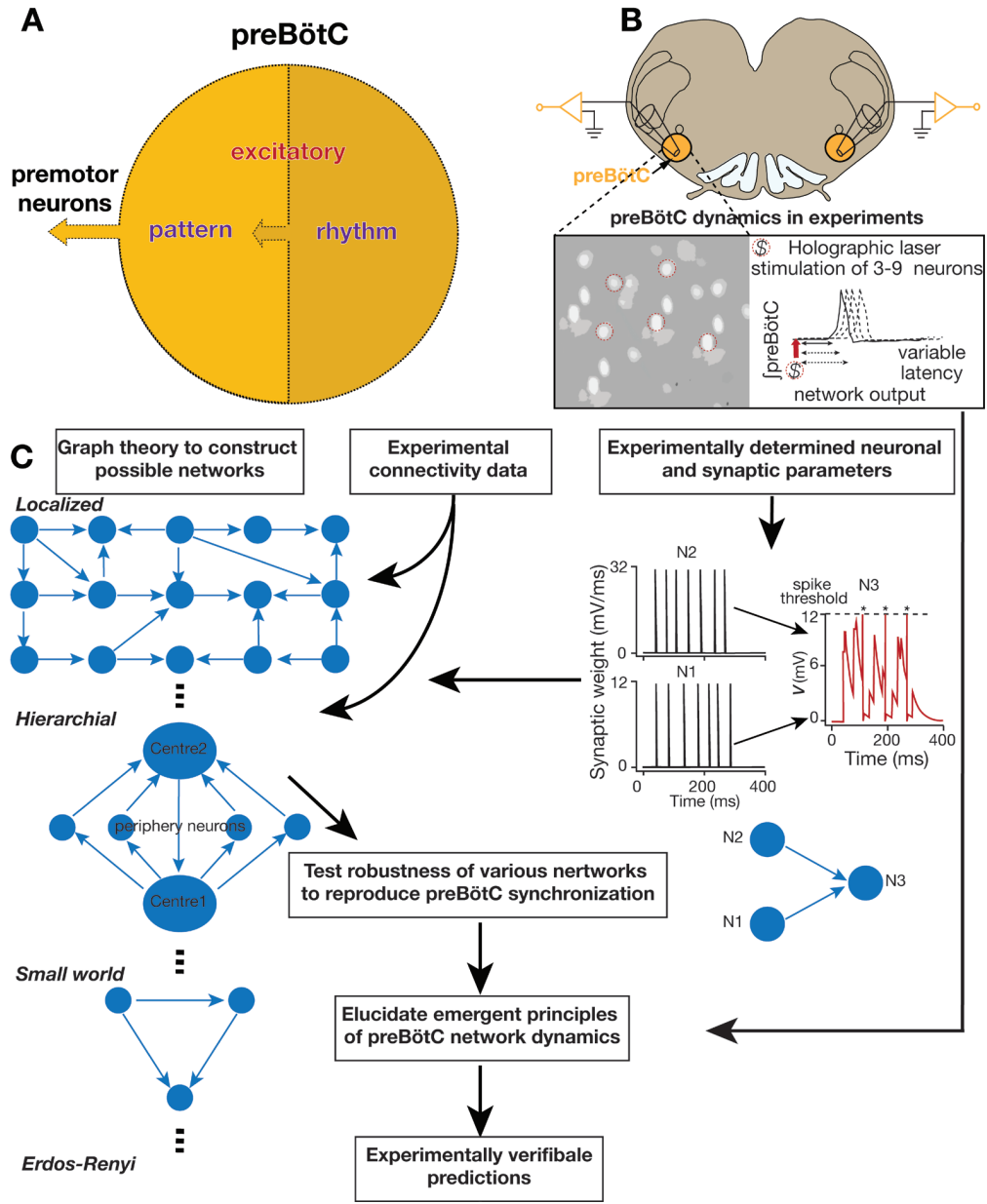

**Figure S1. Flow chart for model construction and testing.** *Related to Figure 2* **A**, Rhythm-generating preBötC neurons project to pattern-generating preBötC excitatory neurons that in turn project to premotor neurons. **B**, holographic photostimulation of 3-9 preBötC inspiratory neurons (out of ~1500 total excitatory neurons) in rhythmic neonatal mouse slices generate bursts at delays ~100-500 ms (Figure 1 F-G). **C**, Various network models tested to determine if they capture preBötC dynamics depicted in **B**: (left) Network topologies as graphs with edges connected through experimentally determined connection probability between putative rhythmogenic neurons (Rekling *et.al.*, 2000)[16]. The nodes were modeled as leaky integrate and fire (LIF) neurons (right) with intrinsic and synaptic properties taken from experiments; (right) synaptic activation from two ("laser") stimulated model rhythmogenic neurons N1 and N2 (presynaptic) that project to neuron N3 (postsynaptic); their activation times represent the arrival of spikes form N1 and N2 at their respective synapses on N3 with weights indicated (left, black traces) consequently, changing in somatic potential of neuron N3 from resting potential (red, right). When the somatic potential increases above spike activation threshold $V^* = 12$ mV ($V \cong -48$ mV), N3 generates an action potential (*), followed by its potential dropping to $V_{rest} = 0$ mV ($V_{rest} \cong -60$ mV) for refractory period of 3 ms.

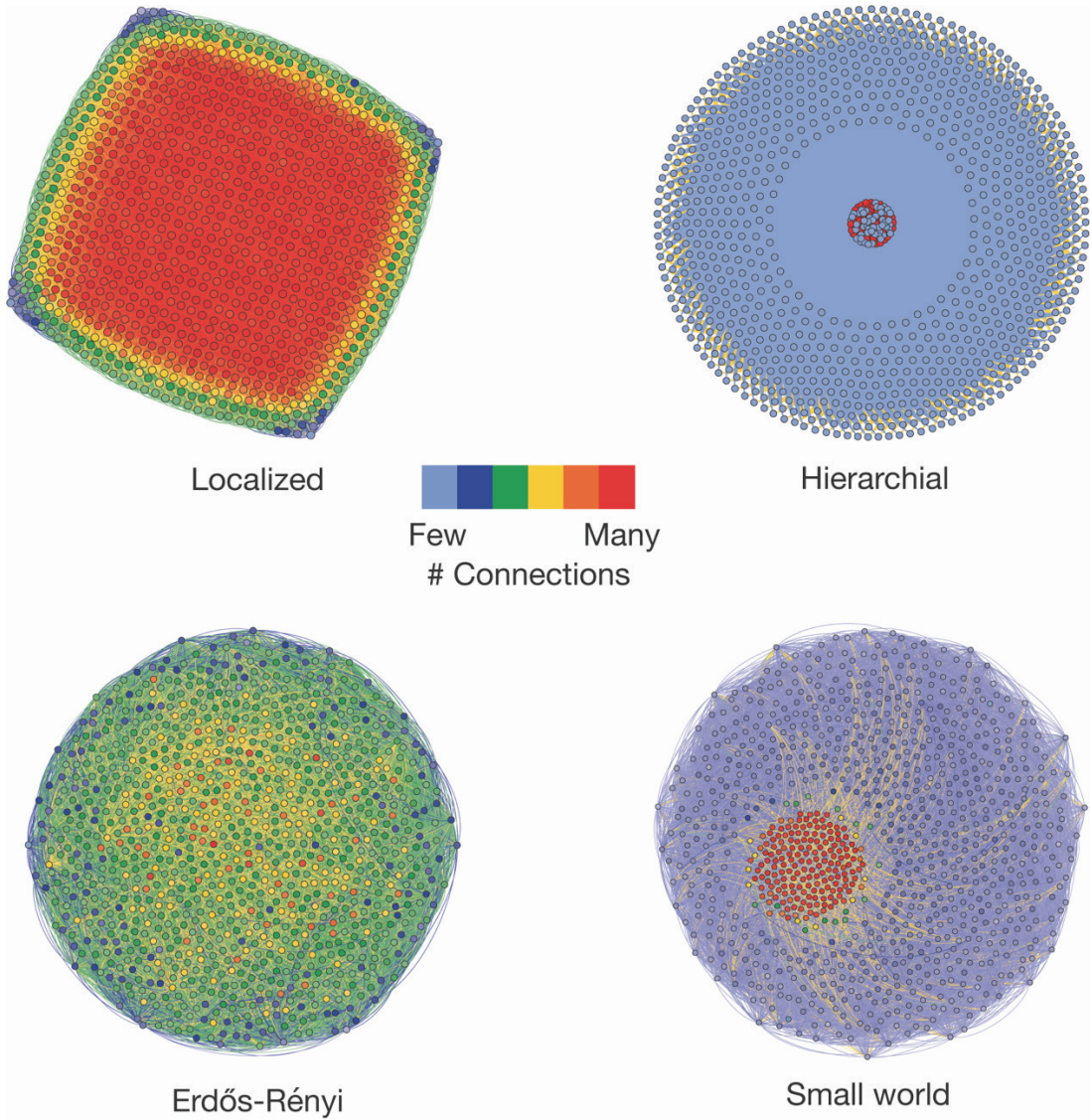

**Figure S2.** *Related to Figure 2*. Various networks represented on a force-based (Fruchterman-Reingold algorithm in the *Gephi* software) layout. linked nodes (neurons) are *pulled* together and unrelated nodes are farther apart; most strongly linked nodes (through direct connections or common inputs) are at the center and the least linked ones at the periphery. Nodes (and their edges) are color coded based on number of their projections, i.e., outward synapses with warmer shades representing more connections. These figures are enlarged versions of the network layouts presented in Figure 2.

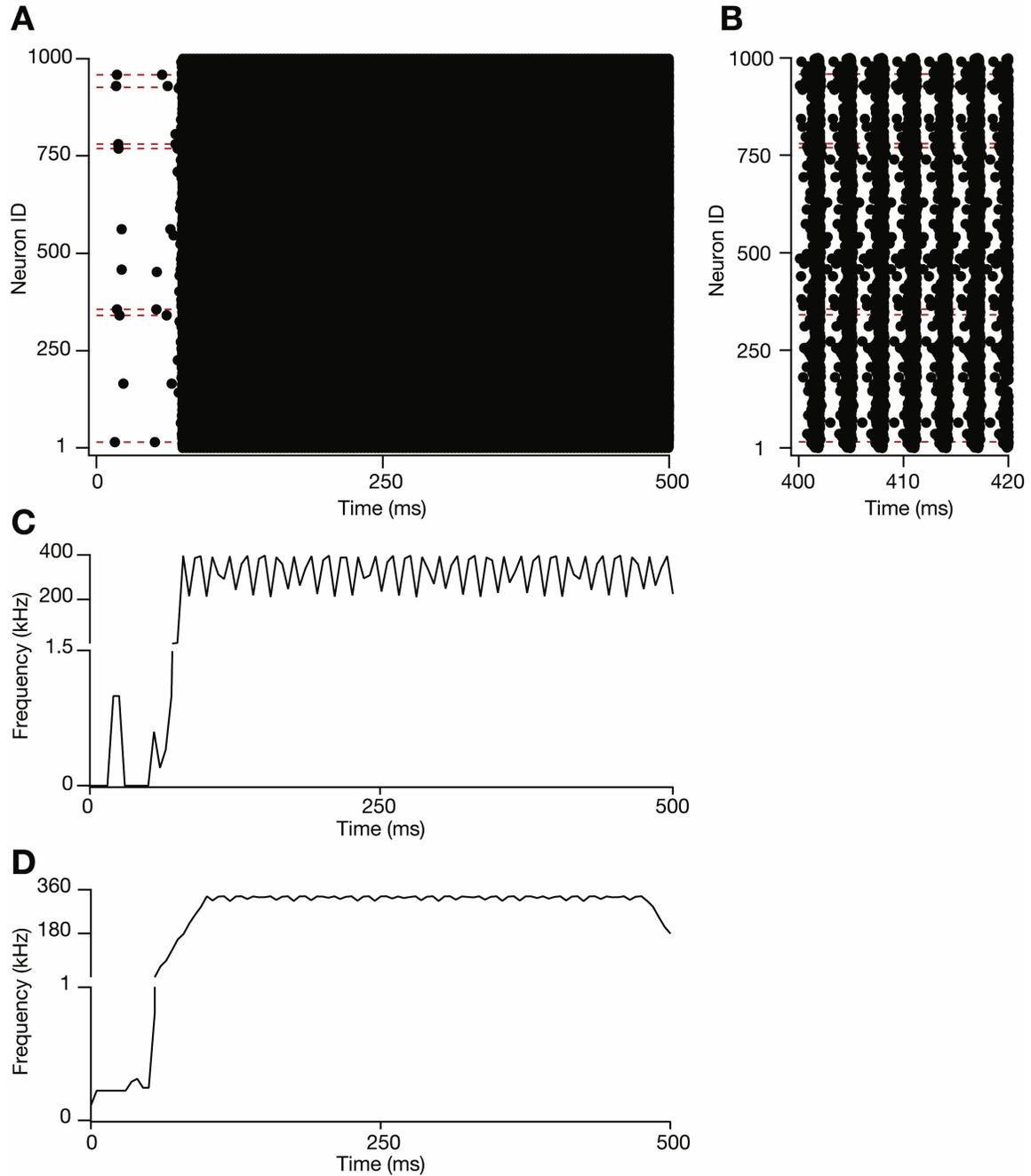

**Figure S3**. *Related to Figure 4*. **A**, Model output when the same set of randomly selected 7 neurons (on the dashed lines) was stimulated to fire seven spikes each at 25 Hz with 5 ms jitter, i.e., SD to model holographic uncaging of glutamate onto these neurons as in (Kam et. al., 2013)[30.] Each dot represents the time of spike (abscissa) from corresponding neuron (ordinate), note that once the network synchronized it continues in the high-frequency firing, i.e., bursting mode even after the stimulated spikes, from the seven neurons, ended at ~300 ms. **B**, Enlarged region from (a) exhibiting firing rate modulation due to the refractory period, of 3 ms, in the model neurons. **C**, average firing rate of network computed by averaging network activity in a moving window of 40 ms with 5 ms step increment.

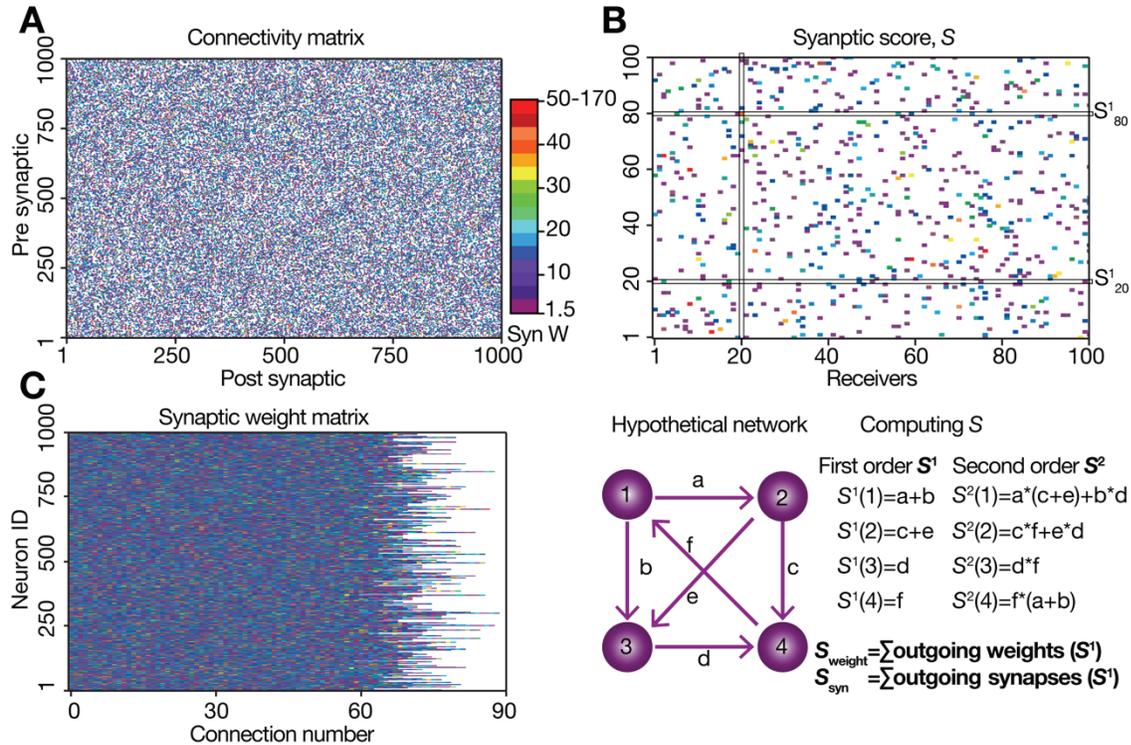

**Figure S4. Computing efferent synaptic score from network topology**. *Related to Figures 5-6*. The efferent synaptic score, *S*, is neuron property defined by its network connectivity. **A,** Connectivity matrix of ER network with lognormal (LN) weight distribution. Presynaptic neurons (ordinate) project to other neurons labelled as postsynaptic (abscissa) with synaptic weights, randomly drawn from a LN distribution. Each dot represents synaptic weight (Syn W; color-coded) of presynaptic #i (1-1000) to postsynaptic #j (1-1000). B, top, expanded section from A (100x100) showing contribution of synaptic connections towards *S*. For example, the first order *S* of neuron #80 ($S^1_{80}$) is the sum of all synaptic weights with which it connects to other neurons. In this example, neuron #20 is postsynaptic to neuron #80, thus, $S^1_{20}$ contributes to the second order *S* of neuron #80 ($S^2_{80}$) (Second order *S* ($S^2$) also accounts for the connectivity of the next neighbors of a given neuron, see text for the calculation of $S^2$); bottom, example of *S* calculation in a hypothetical network. For a given neuron, $S_{weight}$ is the sum of its efferent synaptic weights and $S_{synout}$ is the number of its efferent synaptic connections. **C**, plot of number of outward connections of neurons in ER network with their synaptic weights color-coded as in A and B. This network was used in Figure 5; *S* varies from neuron to neuron as their outward synaptic connections vary, per the ER graph, and as their output synaptic weights vary due to the LN weight distribution.

# Supplemental Tables

## Table S1. Model parameters

| Parameter | Value |
| --- | --- |
| $V_{rest}$ | -60 mV; -65 mV (figure7) |
| V* | -48 mV |
| $\tau_m$ | 25 ms |
| $\tau_s$ | 0.5 ms |
| $\Delta t_{ij}$ | 1.3±1.1 ms |
| $W_{ij}$ | 300±160 (mean ± SD) mV/ms =15±8 in step size (0.05 ms) |
| $\tau_{delay}$ | 20±3 ms |
| $T_{laser}$ | 39±5 ms |
| $n_{spikes}$ | 7 |
| $f_{noise}$ | 0.5-2 Hz |

**Table S2. Prediction accuracy of various generalized *S* quantities for network burst**

| Parameter | Accuracy | Standard deviation |
|---|---|---|
| $\Sigma (W*Y^3)^2$ | 0.73 | 0.01 |
| $\Sigma W*Y^4$ | 0.72 | 0.02 |
| $\Sigma (W*Y^3)^3$ | 0.72 | 0.006 |
| $\Sigma (W*Y^4)^2$ | 0.72 | 0.009 |
| $\Sigma Y^4$ | 0.72 | 0.01 |
| $\Sigma (W*Y^4)^3$ | 0.71 | 0.01 |
| $\Sigma Y^4$ | 0.71 | 0.02 |
| $\Sigma Y^6$ | 0.71 | 0.007 |
| $\Sigma Y^6$ | 0.70 | 0.006 |
| $\Sigma W*Y^3$ | 0.70 | 0.01 |
| $\Sigma Y^3$ | 0.70 | 0.01 |
| $\Sigma Y^3$ | 0.70 | 0.01 |
| $\Sigma (W*Y^2)^3$ | 0.70 | 0.01 |
| $\Sigma Y^9$ | 0.69 | 0.01 |
| $\Sigma Y^8$ | 0.69 | 0.008 |
| $\Sigma Y^{12}$ | 0.68 | 0.01 |
| $\Sigma (W*Y^2)^2$ | 0.68 | 0.009 |
| $\Sigma W*Y^2$ | 0.66 | 0.009 |
| $\Sigma Y^2$ | 0.66 | 0.02 |
| $\Sigma Y^2$ | 0.66 | 0.01 |
| $\Sigma (W*Y)^3$ | 0.59 | 0.008 |
| $\Sigma Y$ | 0.57 | 0.02 |
| $\Sigma (W*Y)^2$ | 0.57 | 0.01 |
| $\Sigma W*Y$ | 0.56 | 0.01 |
| *All above simultaneously* | 0.74 | 0.006 |